\documentclass[sensors,article,accept,moreauthors,pdftex,10pt,a4paper]{Definitions/mdpi}
\firstpage{1}
\articlenumber{1415}
\pubvolume{19}
\issuenum{6}
\pubyear{2019}
\copyrightyear{2019}
\history{Received: 8 February 2019; Accepted: 18 March 2019; Published: 22 March 2019}
\updates{yes}

\Title{The Life of a New York City Noise Sensor Network}

\Author{Charlie Mydlarz
$^{1,2,*}$ \href{https://orcid.org/0000-0001-7061-0638}{\orcidicon}, Mohit Sharma $^{2}$, Yitzchak Lockerman $^{3}$, Ben Steers $^{2}$ and Claudio Silva $^{3}$ and~Juan~Pablo~Bello~$^{1,2}$}

\AuthorNames{Charlie Mydlarz, Mohit Sharma, Yitzchak Lockerman, Ben Steers, Claudio Silva, Juan Pablo Bello}

\address{%
$^{1}$ \quad Music and Audio Research Laboratory, New York University, New York, NY 10012,
USA; jpbello@nyu.edu\\
$^{2}$ \quad Center for Urban Science and Progress, New York University, Brooklyn, NY 11201, USA; mohit.sharma@nyu.edu (M.S.); bsteers@nyu.edu (B.S.)\\
$^{3}$ \quad Tandon School of Engineering, New York University, Brooklyn, NY 11201, USA; yitzchak.lockerman@nyu.edu (Y.L.); csilva@nyu.edu (C.S.)}

\corres{Correspondence: cmydlarz@nyu.edu; Tel.: +1-646-997-0536}

\abstract{Noise pollution is one of the topmost quality of life issues for urban residents in the United States. Continued exposure to high levels of noise has proven effects on health, including acute effects such as sleep disruption, and long-term effects such as hypertension, heart disease, and hearing loss. To investigate and ultimately aid in the mitigation of urban noise, a network of 55 sensor nodes has been deployed across New York City for over two years, collecting sound pressure level (SPL) and audio data. This network has cumulatively amassed over 75~years of calibrated, high-resolution SPL measurements and 35~years of audio data. In addition, high frequency telemetry data have been collected that provides an indication of a sensors' health. These telemetry data were analyzed over an 18-month period across 31 of the sensors. It has been used to develop a prototype model for pre-failure detection which has the ability to identify sensors in a prefail state 69.1\% of the time. The~entire network infrastructure is outlined, including the operation of the sensors, followed by an analysis of its data yield and the development of the fault detection approach and the future system integration plans for this.}


\begin{document}

\section{Introduction\label{sec:intro}}
Noise pollution is an increasing threat to the well-being and public health of city inhabitants \cite{isling2004}. It has been estimated that around 90\% of New York City (NYC) residents are exposed to noise levels exceeding the Environmental Protection Agency's (EPA) guidelines on levels considered harmful to people \cite{hammer2014environmental}. This exposure has proven health effects such as: sleep disruption, hearing loss, hypertension, and heart disease \cite{bronzaft2007neighborhood, hammer2014environmental, fritschi2012introduction}. In addition, studies have revealed the negative impacts of noise on the learning of children. This cognitive impairment includes effects such as decreased memory, reading skills and lower test scores \cite{bronzaft2007neighborhood, basner2014auditory}. The mitigation of this noise pollution is far from trivial and the difficulty of citywide enforcement is exacerbated by the transient nature of the various noise sources and the people hours required to effectively monitor and intervene.


At the Sounds Of New York City or SONYC project, we have deployed a network of over 55 low-cost acoustic sensor nodes across NYC to facilitate the continuous, real-time, accurate and source-specific monitoring of urban noise \cite{Bello:SONYC:CACM:18}. This enables large-scale analysis of urban noise activity including predictive noise impact models and interactive 3D visualizations to reveal noise patterns across space and time. These tools have been used to drive a data-driven approach to noise mitigation by feeding relevant, actionable, and timely data to those tasked with reducing noise. A number of the sensor nodes have been operational since May of 2016, resulting in the accumulation of vast amounts of calibrated sound pressure level (SPL) data and its associated metrics. Cumulatively to date, 75~years of SPL and 35~years of raw audio data has been collected from the sensor network. This data can be used to identify longitudinal patterns and often overlooked occurrences of noise pollution across urban settings \cite{Miranda:TimeLattice:CGF:18,mydlarz_internoise_2017}.


In New York City (NYC), noise is monitored via its 311 civil complaint service \cite{nyc_311}, which has logged over 3~million noise related complaints to date, outnumbering any other complaint type. The service is effectively a crowd-sourced means of urban noise monitoring, but the data that it generates has many shortcomings that limit its validity as an indicator of the actual noise conditions of the city. Firstly, it is sparse in space and time and secondly it is biased by noise source type and socio-demographic factors, as more affluent Manhattan neighborhoods are more than twice as likely to report noise issues when compared to others that could be considered under-served~\cite{EpiBrief2014}. These complaints are also handled in a reactive fashion with a critical mass in a specific area resulting in noise enforcement action by city agencies. This approach relies on a build-up of negative human response to an objective and measurable stimulus, something that an acoustic sensor network is particularly suited to monitoring. Noise source trends can be tracked over time with the insight required for interventions provided at an earlier stage, prior to a large number of people filing complaints. A sensor network's ability to continuously monitor allows for the creation of complimentary data to crowd-sourced approaches due to its high temporal resolution and longitudinal nature. Manual long-term noise monitoring is rarely carried out due to its cost and expert personnel requirements. In addition, deployments in under-served neighborhoods provide an objective and unbiased indication of noise conditions, overcoming the majority of the limitations of 311. The combination of crowd-sourced noise complaint data and continuous objective measurements provide a far more actionable and insightful data-set for the people carrying out interventions with the goal of improving urban noise conditions for city~inhabitants.

In this paper: (1) We'll review prior work on acoustic sensor networks for noise monitoring, and show that these approaches either lack data quality, scale and/or longevity; or are expensive commercial ventures for which information is not forthcoming. (2) We propose a new sensor infrastructure that is high quality, large-scale, and durable, but also research-focused and open, therefore providing a unique opportunity for new knowledge of value to the sensors community. We~give a detailed account of the network's design, implementation, deployment and operation. (3)~Given the longevity of the network, we provide a detailed historical analysis of network performance in real-world conditions to highlight challenges and opportunities. (4) We conclude by discussing the conclusions and future directions for the network.




\section{Prior Work}





Accurately measuring environmental noise data across a number of locations for long periods of time under varying conditions is a difficult task. Effective urban noise monitoring requires the collection of accurate longitudinal acoustic data from a large number of static locations that includes sound pressure level (SPL) metrics and raw audio data. The literature has many examples of these sensor networks with a broad range of: network sizes, accuracies, capabilities, price points, and~deployment durations. Another attribute that will be discussed is the accessibility or openness of the data generated by these sensor networks, which can increase their utility and general benefit. The~following examples of prior work will be reviewed with a focus on these attributes.

Since the mid 2000s, consumer smartphone technology has allowed for in situ sound pressure level (SPL) measurements to be crowdsourced for applications such as citizen science led noise mapping projects \cite{murphy2016smartphone, kardous2014evaluation, kuhn2013analyzing, roberts2016improving, maisonneuve2010participatory, kanjo_2010, guillaume2016noise, li2018participatory, picaut2019open}. This approach is frequently referred to as \textit{participatory sensing}. Whilst~the ability to gather data from more diverse locations is unmatched by other approaches, firstly, their~spatial-temporal coverage is less than adequate, and, secondly, their accuracy is questionable due to the lack of calibration and the variation in microphone sensitivity between device models and even device condition. The penetration demographics of smartphone users in the US is also biased towards well paid individuals with a college level education \cite{smartphone_penetration_2018}. Interestingly, this bias is also observed in the propensity of individuals to file a complaint to the New York City (NYC) 311 civil complaints service, with a disproportionate amount arriving from the higher median income Manhattan, despite survey data showing that reported noise disruption prevalence is similar across Manhattan, Brooklyn and the Bronx \cite{EpiBrief2014}.

A static noise monitoring network from Piper et al. in 2017 \cite{piper2017construction} focuses on the data collection from a network of 16 low-cost nodes across a six~month deployment period. These sensors used custom designed microphones that were able to generate high quality acoustic data. One minute resolution noise data was collected at a construction site. The availability or uptime of the sensor network is discussed with favorable values of 67.6\% to 97.8\%. Power failures and adverse weather conditions were seen to have an effect on uptime. The shortcomings to this study were mainly its relatively short duration and limited scale; however, they were able to generate some useful inference on the noise conditions of the localized site, proving the efficacy of longitudinal noise sensor network deployments for the understanding of noise patterns. This project also releases its data via online dashboards to relevant stakeholders, making its accessibility for wider use one of its key benefits, although these dashboards are not available to the public.

The Array of Things Project \cite{Catlett:2017:ATS:3063386.3063771} focuses on the use of a static sensor network to monitor a range of environmental parameters including noise levels in Chicago. This network currently stands at 100 deployed nodes, most of which have been operational for over six~months. This ambitious and longitudinal project aims to deploy hundreds more nodes over the coming years. This scale of deployment will surely generate large amounts of urban noise data; however, the accuracy of the project's noise sensing hardware could be considered low as the microphone itself is mounted within the sensor housing causing acoustic resonance and filtering effects. Its cost per sensor unit is high at the \$2000~USD mark; however, a strategic partnership with the Chicago Department Of Transport (DOT) means that maintenance visits are facilitated by their engineers, reducing upkeep costs over the networks lifetime. The data generated by the project is also easily accessible online to the public in real time, a welcome rarity for this type of research institution led network.

Sensor networks that excel in the generation of accurate SPL data compliant with the precision class/type~I IEC~61672 accuracy standards \cite{international2013electroacoustics} are the commercial solutions offered by companies such as: Bruel and Kjaer (B\&K), Norsonic, and 01dB. For example, B\&K's Noise Sentinel system \cite{bk_noise_sentinel_web} provides highly accurate SPL measurements, wireless data transmission, simple threshold based event detection/audio recording, and a private web portal for real-time data visualization. This network is typically deployed over a number of months for specific projects, including construction and mining projects to monitor noise levels and feedback the data in real time to stakeholders. One of their more longitudinal deployments is around the New York Metropolitan area, monitoring aircraft noise and their compliance with local air noise codes. This deployment (1 of 50 globally) has been ongoing for over four~years and consists of 37 precision noise sensors continuously monitoring SPL, with real-time data presented on a private web interface \cite{bk_webtrak_portal} and a service level agreement (SLA) of \textgreater95\% uptime of the sensor network. In addition to the logging of SPL data, these sensors record and transmit audio snippets of aircraft fly-overs when a given level threshold is exceeded for manual offline aircraft type classification. This narrowly focused network provides simple but highly accurate longitudinal noise data over a moderately large area; however, the yearly cost per sensor node comes in at \$37,000~USD, which results in over \$1.35M~USD/year for the New York Metropolitan area's deployment. This high cost limits the scale of deployments to focused studies and is typically reserved for large budgets as is the case in this example. These networks also only generate noise SPL data, and require costly manual labelling of audio data to identify the offending sources.

A research project that also makes use of commercial environmental monitoring stations is the SmartSensPort-Palma project \cite{pabi_2018} in Mallorca, Spain. Their network of eight sensors around a commercial shipping area of Mallorca measured environmental pollution, including noise data over an eight month period at the general use IEC~61672 class/type~II level of accuracy. Their sensors are made by the Spanish remote sensing company Libelium \cite{libelium_2019} and retail for upwards of \$1800~USD each, with a total deployed network cost of \$51,000~USD. Their aim was to uncover relationships between port activity and environmental pollutants. This relationship wasn't uncovered, but their conclusion was that longer study periods with larger numbers of sensors was required to adequately model environmental noise patterns to generate actionable data for stakeholders. Whilst~orders of magnitude lower in cost than the previous B\&K network, the cost per sensor node is still relatively high, limiting~deployment scale. An issue that arose was urban traffic noise dominating the signal measured at the eight sensor locations rather than the desired port noise. The Libelium sensor's low-power computing core only allows for basic SPL measurements rather than more advanced signal processing that would have allowed for this distinction at the network edge. Open access to this data is not apparent from the project's documentation.

A noise monitoring network with a broader set of proposed applications was created by the Bruitparif organization, called the \textit{Rumeur Network} \cite{mietlicki2015}. Bruitparif was formed in 2004 by the French Ministry of Environment to monitor environmental noise in France. Their network consists of long-term noise monitoring stations at fixed locations for longitudinal noise studies (many years), alongside medium-term deployments (weeks to years) to evaluate the impact of noise producing facilities, and~short-term deployments (hours to days) of mobile sensors to study short term noise events. They~partner with European universities to study the effects of noise on various actors within urban settings including the results of enhanced mitigation strategies. The sensors that make up this network consist of 50 \$2500~USD monitoring stations with multi-core computing capabilities deployed for long-term periods, gathering high quality audio and accurate acoustic data at the IEC~61672 class/type~I level, including raw audio collection. These are complemented by 350 \$550~USD lower-cost devices with shorter deployment durations that transmit simple SPL values at the IEC~61672 class/type~II accuracy level. The high processing power available within these more advanced monitoring stations provide a stream of raw audio recordings alongside the SPL data when SPL thresholds are exceeded, providing a richer understanding of the localized noise environment over long periods of time. Their~deployment duration is also impressive with a core set of these advanced sensors operational for over six years. Whilst this project has accumulated large amounts of accurate longitudinal SPL data from a large geographical area, the context on the generating sources behind this data is limited, making it difficult to draw complete inferences on their urban noise situation. Their data is also accessible via a real-time public facing web interface.

A similar noise monitoring network to the Rumeur Network is the \textit{DYNAMAP Life+ Project} \cite{dynamap2018}, where 93 noise sensors were deployed across Rome and Milan for periods of several months to aid in the production of accurate noise maps for the cities. This network as in the case of Rumeur used a hybrid approach to sensing with the use of high and low capability sensor nodes. These high capacity nodes provide more advanced capabilities beyond simple SPL measurements, such as anomalous source event detection to increase the accuracy of road traffic noise SPL measurements. These networks were able to produce noise data at 30~s intervals which resulted in more accurate noise maps than was traditionally possible using mainly prediction based sound propagation models. However, the SPL data accuracy of this system is variable due to the use of low cost off-the-shelf microphones across the hybrid network of sensors. Whilst the hybrid approach to sensing has its benefits in terms of expanded battery/solar powered deployments using the low capacity nodes, the consistency between measurements of the high and low capability nodes is unclear and possibly significant.

Other relevant projects include the MESSAGE project \cite{bell2013novel} in the UK and the IDEA project \cite{botteldooren2013internet} in Belgium. The MESSAGE project deployed up to 100 low-cost noise sensors with limited dynamic range and a relatively low temporal rate of noise measurements due to their battery powered operation. Whilst large in scale, the data accuracy is questionable for more focused noise monitoring aside from general road traffic noise estimations, with shorter duration deployments detailed in their publications. The IDEA project utilises a more accurate sensor system, but lacks the longevity of other studies.

To summarize these, Table~\ref{tab:network_features} shows the key attributes for each aforementioned study: noise data quality, network scale, study longevity, system affordability, and its data accessibility. Included are the attributes for the SONYC noise sensor network. This tabulation is intended to help emphasise the contribution to the literature that this paper will provide. In summary, whilst each of these given studies of sensor networks has their strengths, none combine the key attributes and exploration of: large network size, accuracy, low-cost, extended deployment durations, and accessibility of generated data. This paper aims to provide findings and insight from a low-cost sensor network focused on accurate noise data collection that has been operational for a long period of time over a large urban area. The~following sections detail the SONYC sensor network's deployment, design, and~continued~operation.
\begin{table}[H]
\centering
\caption{Existing sensor network literature summary. \label{tab:network_features}}
\begin{tabular}{cccccc}
\toprule
& \bf{Data Quality} & \bf{Scale}    &  \bf{Longevity}   & \bf{Affordability}    & \bf{Accessibility}\\
\midrule
Participatory sensing       & low               & large         & short             & high                  & low\\
Piper et al.                & high              & small         & short             & high                  & low\\
Array of Things             & low               & large         & long              & high                  & high\\
B \& K/Norsonic/01dB        & high              & small         & long              & low                   & low\\
SmartSensPort-Palma         & high              & small         & short             & low                   & low\\
Rumeur                      & high              & large         & long              & low                   & high\\
Dynamap Life+               & low               & large         & short             & high                  & high\\
MESSAGE project             & low               & large         & short             & high                  & low\\
IDEA project                & high              & large         & short             & high                  & high\\
SONYC                       & high              & large         & long              & high                  & high\\
\bottomrule
\end{tabular}
\end{table}

\section{Sensor Network Deployment and Data Collected}

The SONYC sensor network consists of 55 remote acoustic sensors or nodes deployed across New~York City (NYC), each containing a quad core single board computer (detailed in Section~\ref{sec:sensor_node}). The~first sensor was deployed in May of 2016 so data collection has been ongoing for close to three years as of writing. The deployment considerations of the sensor network are discussed in Section~\ref{sec:deployment}, followed by a description of the data that they generate in Section~\ref{sec:data_collected}.



\subsection{Deployment}
\label{sec:deployment}
\begin{figure}[H]
\begin{center}
\includegraphics[width=0.50\textwidth]{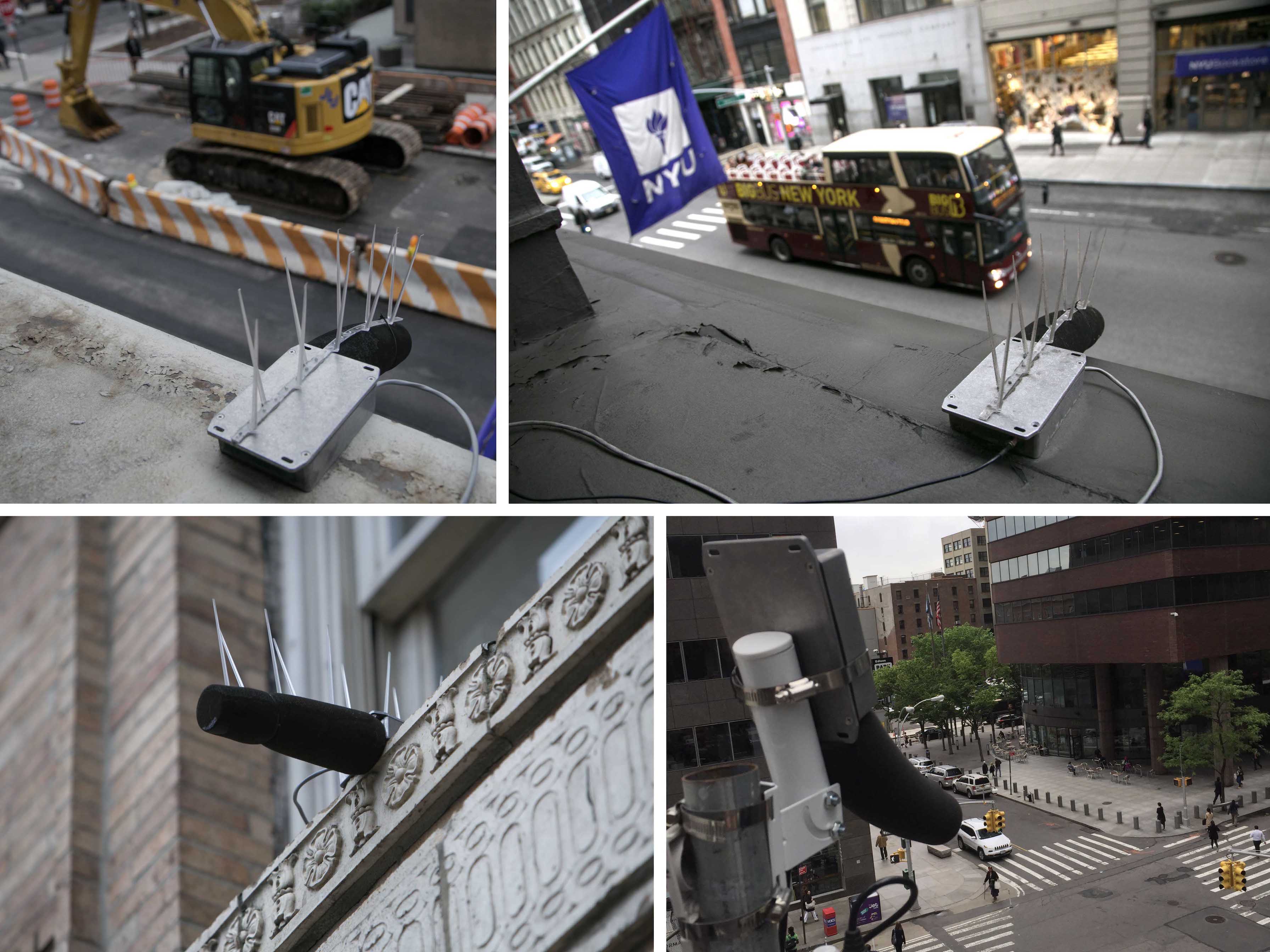}
\caption{Acoustic sensor nodes deployed on New York City streets.}
\label{fig:node_deployed}
\end{center}
\end{figure}

The physical deployment of the sensor nodes is typically surface mounted using industrial adhesive to a horizontal or vertical surface such as a window ledge, wall, or pole mounted. Figure~\ref{fig:node_deployed} shows a selection of these deployed sensor nodes. Deployments are always external and efforts are taken to mount the nodes away from existing sources of noise at the building edge---for example, air conditioning units or vents. Care is also taken to ensure the node has an un-occluded view of the street and is as far away from main exterior walls as possible to reduce the artificial boosting of measured SPL levels when too close to large hard surfaces.

\begin{figure}[H]
\begin{center}
\includegraphics[width=0.6\textwidth]{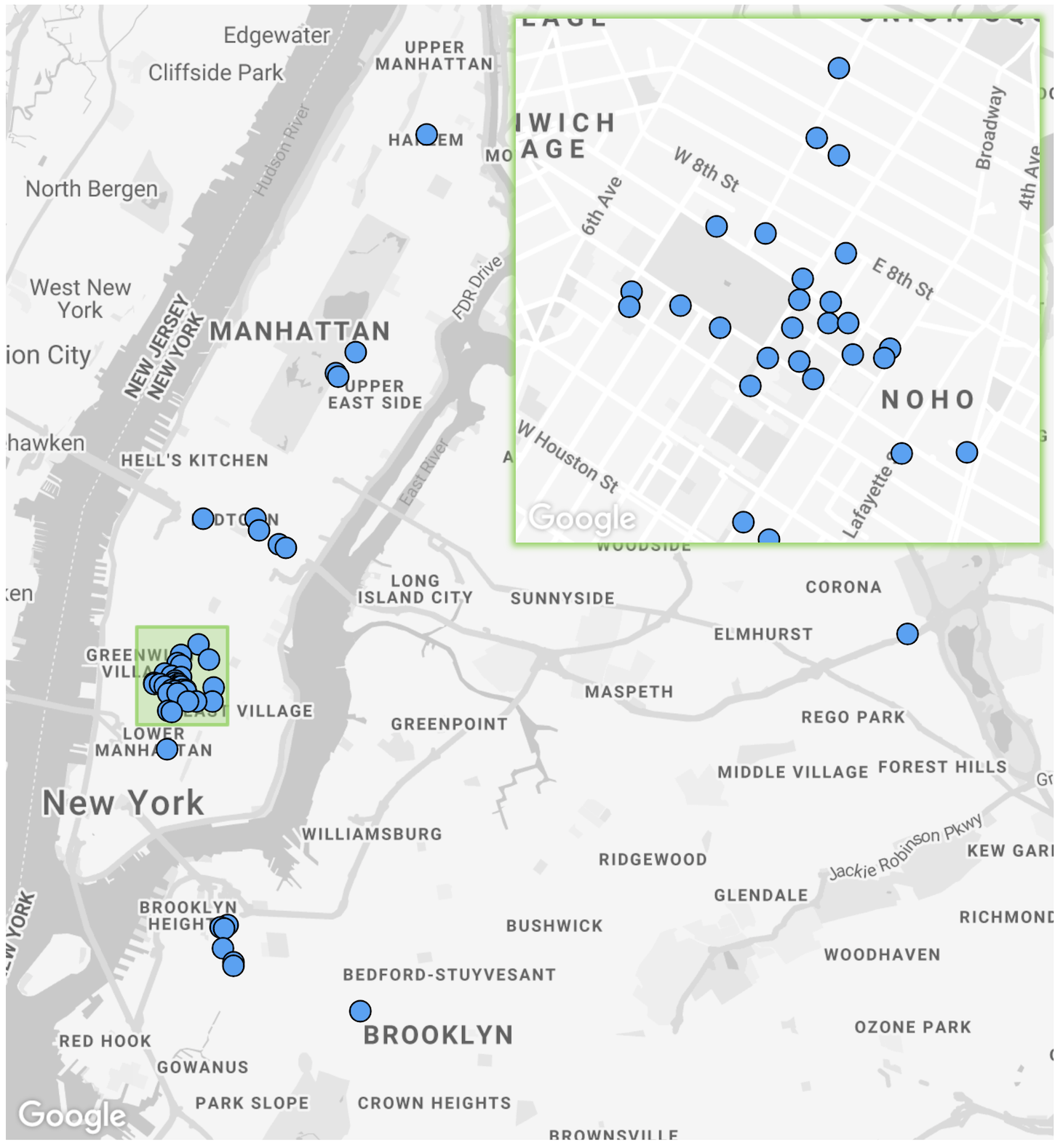}
\caption{Sensor network deployment map with an entire network of 55 sensors and a green focus area of dense Washington Square Park deployment.}
\label{fig:sensor_map}
\end{center}
\end{figure}

The current sensor locations can be seen in Figure~\ref{fig:sensor_map}. A general rule of at least a one-block distance between sensor nodes was adhered to, unless there was a particular point of interest close by, such as a long term construction project or major roadway. The initial nodes were deployed on NYU locations and are densely clustered around the Washington Square Park area in Manhattan, providing a test-bed for the focused analysis of localized noise issues, such as the predominance of after hours construction noise complaints \cite{mydlarz_internoise_2017}. A number of deployments were carried out in partnership with local business improvement districts (BIDs). These deployments made use of bucket lifts to mount nodes on BID owned and maintained light poles, usually greater than 15~feet from the ground. These types of deployment are difficult to access for maintenance so measures have to be taken to ensure continuous uptime and the ability to remotely troubleshoot issues. Some of these measures have been described in Section~\ref{sec:op_func}, such as the \textit{sonyc-lifeline} strategy for close proximity node connection and repair in the event of a software issue.

\unskip
\vspace{-6pt}
The variation in deployment characteristics such as sensor height and flat surface/pole based mounting will likely produce variable SPL measurements between deployment types. This makes the sensors more adept at measuring decibel change over time rather than providing directly comparable absolute SPL values across different deployment types. The majority of the sensors are mounted onto window ledges at heights ranging between 15--25~feet. In two locations, sensors are mounted on the 6th floor of a building at around 60~feet above the ground. A smaller number are mounted on tubular street light poles at heights between 15--25~feet. In practice, over long deployment durations the height difference between sensor nodes would result in negligible SPL variations between the sensors ranging 15--35~feet in height, as the most significant difference in absolute SPL measurement between different sensors occurs when a noise source is directly below the sensor for long periods of time. This event is less likely than noise sources occurring at some distance from the sensor where the distance to the sensor becomes similar when varying the sensors height from the ground. However, making~absolute comparisons of SPL values between different sensor locations is prone to error. Comparative measurements averaged over extended durations of time are less prone to this error.

The primary limits on deployment locations are the node's reliance on a wired power connection and the required proximity to a Wi-Fi access point. To open up more potential deployment locations across NYC, a partnership was developed with a major provider of public Wi-Fi. This partnership enabled deployments in more diverse locations where their Wi-Fi networks were broadcast. With~this requirement satisfied, it was easier to successfully approach other partners who could provide deployment locations with mounting possibilities and power. With reduced data transmission requirements, the possibility of cellular connectivity becomes viable, allowing for enhanced deployment ranges and locations. This is work in progress and will be discussed in Section~\ref{sec:future_work}.

\begin{figure}[H]
\begin{center}
\includegraphics[width=0.4\textwidth]{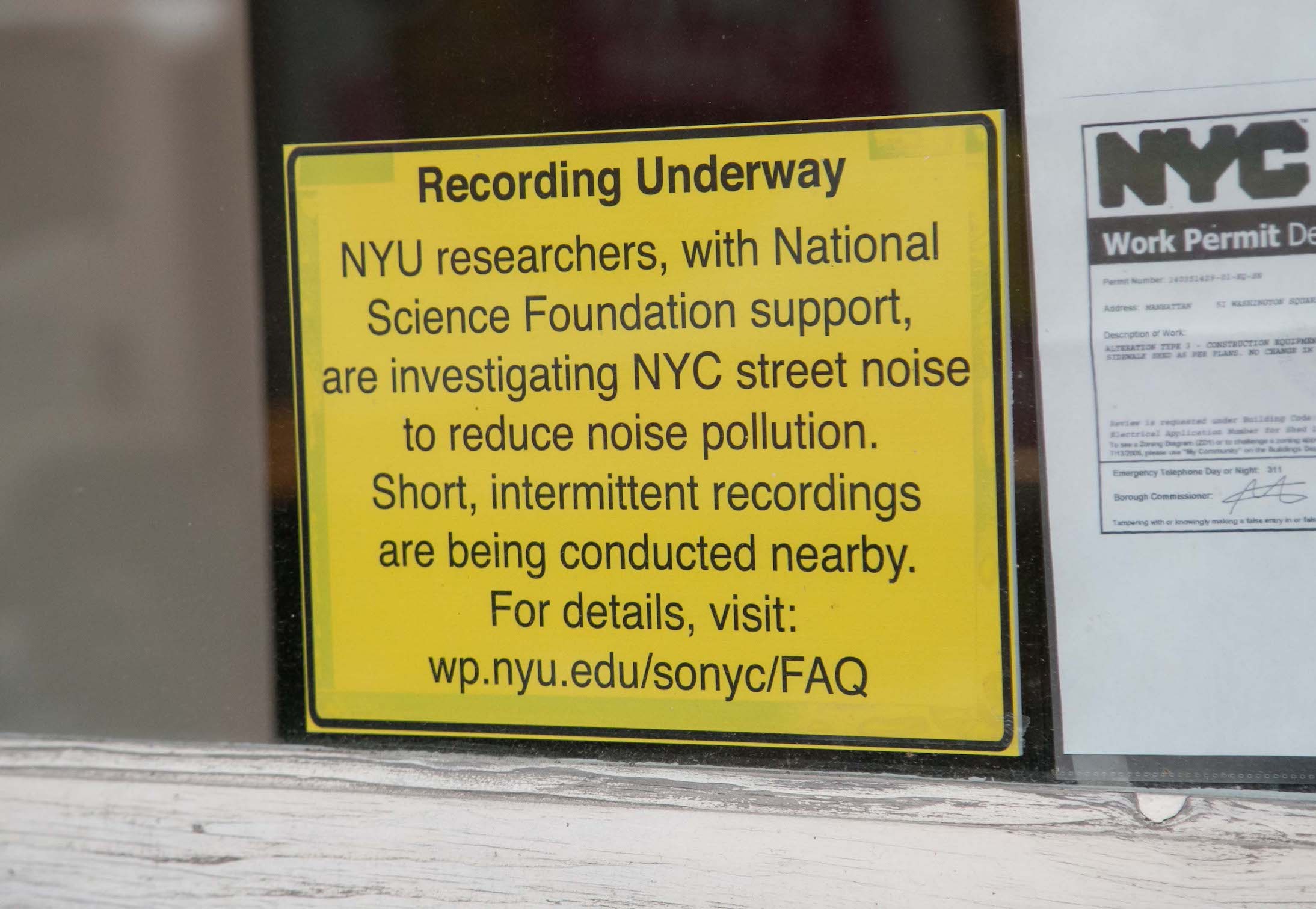}
\caption{Street-level signage mounted in close proximity to sensor node location for public information.}
\label{fig:deployment_sign}
\end{center}
\end{figure}

An often underestimated consideration in deploying our sensor nodes is the requirement for a highly visible street-level sign briefly outlining the research that is going on. This must be located close to the deployed node so passersby are aware that short snippets of audio data (discussed~in~Section~\ref{sec:data_collected}) are being recorded and provides them with a link to the project's frequently asked question (FAQ) site \cite{sonyc_faq} for more information. Permissions to deploy these signs are typically harder to get, as they are public facing and more conspicuous than the sensors. However, the importance of these signs cannot be overstated. The upfront transparency of displaying the signage can go a long way to alleviate privacy concerns that city inhabitants may have as it provides accessible information on the motivations behind the project, its goals, and steps taken to mitigate these concerns. Great lengths have been taken to address these concerns, such as the submission of sample audio snippets with speech present for review by independent acoustical consultants, who judged them to be unintelligible as conversation. In addition to this, New~York University's (NYU) Institutional Review Board determined that the project's sensing activities are exempt from further human subjects' protection scrutiny. NYU’s Institutional Review Board reviews all proposed research involving human subjects to ensure that any subjects' rights and welfare are adequately protected in accordance with regulations issued by the U.S. Department of Health and Human Services Office for Human Research Protections. Figure~\ref{fig:deployment_sign} shows one of these window mounted~signs.

\vspace{-6pt}
More recent deployments have been primarily driven by the existence of localized sources of high-impact noise such as large building constructions, raised subway lines, or extensive road works as identified by the proliferation of noise complaints in the area. The time and effort required to secure permission to deploy a node with the relevant owner of the infrastructure you are mounting on means that priority must be given to high-impact sites where the node will be providing the most value to the research and/or project stakeholders. The time and money required to build sensor nodes can be outweighed by the time and cost required to gain the necessary permissions to deploy and then maintain these sensor nodes as the network expands over time. A low starting cost of sensor hardware is therefore important to the successful implementation of a scalable sensor network. In~Section~\ref{sec:monitoring}, we~discuss the process of sensor network monitoring and how this aids in the process of network~upkeep.

\subsection{Data Collected}
\label{sec:data_collected}
A sensor node collects three main types of data, in order of priority: sound pressure level (SPL) values, encrypted 10~s audio snippets, and node status data. SPL data is in the form of two comma delimited files within a tar file for 1~s and 0.125~s resolution A, C and un-weighted SPL data generated every minute. These files also contain single and third octave band SPL values to determine acoustic energy at discrete points across the audible spectrum. The encrypted audio snippets are losslessly compressed 10~s audio files contained within gzipped tar files. These audio files are collected for the purposes of machine learning and act as the training data for subsequent automatic noise source identification models, discussed in Section~\ref{sec:future_work}. The node status data is uploaded as a JSON payload with no compression due to its small size. These data types are summarized in Table~\ref{tab:main_data_types}.

\begin{table}[H]
\centering
\caption{Collected data type summary ordered by descending priority including total collected size to~date.\label{tab:main_data_types}}
\begin{tabular}{cccccc}
\toprule
\bf{Description}           & \bf{Format} & \bf{Size} &  \bf{Frequency} & \bf{Cached} & \bf{Total Size}\\
\midrule
Sound pressure level (SPL)  & tar       & 150~KB & 60~s    & yes  & 2~TB\\
Audio snippets              & tar.gz    & 500~KB & 20~s   & yes   & 40~TB\\
Node status                 & JSON      & 1~KB   & 3~s   & no     & 80~GB\\
\bottomrule
\end{tabular}
\end{table}

SPL data is captured continuously and is cached locally in case of upload failures. Audio data are captured and encrypted around three times per minute with randomly spaced intervals between recordings to ensure discontinuity in recordings for privacy reasons. These are also cached locally for persistence if upload fails. Node status data are uploaded every 3~s; however, if an upload fails, the data are not cached, it is refreshed with new data and the upload cycle is repeated after another 3~s. This data are of low enough priority that local disk space is not utilized for it. To date, the sensor network has cumulatively collected 75~years of SPL data and 37~years of audio snippets. The following sections describe the sensor's design and operation that enable it to continuously generate this data.

\section{Sensor Node}
\label{sec:sensor_node}
This section details the hardware and software components that constitute the current version of the acoustic sensor node, including the data collection and networking strategies employed.
\subsection{Sensor Core}
\begin{figure}[H]
\begin{center}
\includegraphics[width=0.75\textwidth]{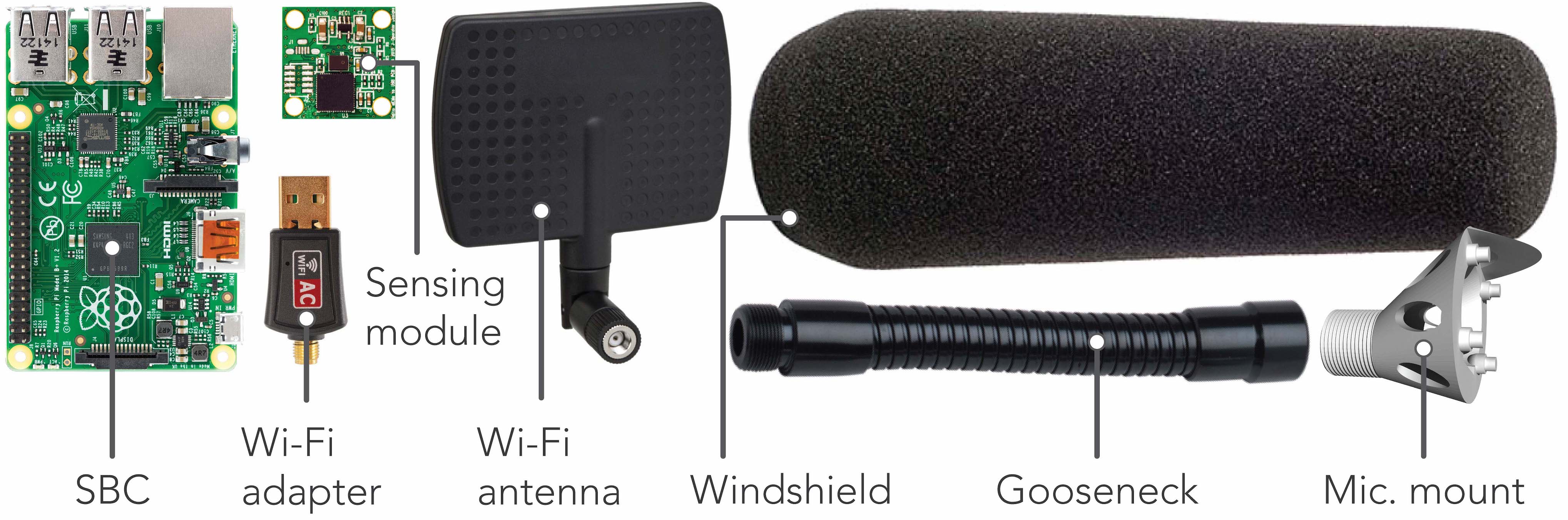}
\caption{Main part list for complete node at relative scale (excluding housing).}
\label{fig:main_components}
\end{center}
\end{figure}

The acoustic sensor nodes primarily consist of off-the-shelf hardware to drive down the overall \$80~USD parts cost of each node. Figure~\ref{fig:main_components} shows the core node components (shown at relative scale to each other). The popular Raspberry Pi 2B single-board-computer (SBC) sits at the core of the node running the Linux Debian based Raspbian operating system, providing all main data processing, collection and transmission functionality. The choice of the Raspberry Pi over the plethora of other SBC choices is mainly due to the maturity and thus stability of the Raspbian operating system and the large online community that has been developed over the many years of the Raspberry Pi's existence. The majority of nodes make use of a 2.4/5~GHz 802.11~b/g/n USB Wi-Fi adapter for internet connectivity; however, a number of nodes also employ a low-cost power-over-ethernet (POE) module which provides internet connectivity and power over a single ethernet cable. Initially, only 2.4~GHz Wi-Fi modules were used, but in some locations with high levels of ambient 2.4~GHz traffic, such as those close to high-rise residential buildings, connectivity was intermittent. This prompted the use of dual-band 2.4/5~GHz USB Wi-Fi modules to allow for the use of the less congested 5~GHz bands if needed. To further enhance Wi-Fi signal strength, we make use of directional antennas. These are more sensitive on-axis, so they can be pointed at the nearest Wi-Fi access point, increasing signal strength and helping to reduce the negative effects of unwanted ambient radio frequency (RF) signals. The~aluminum housing of the sensor provides further radio frequency interference (RFI) shielding and resistance to solar heat gain. In~some locations, bird spikes provide protection from damaging droppings on the node and reduces the chance of birds picking at the microphone windshield. The physical deployment of the sensor nodes is discussed in Section~\ref{sec:deployment}.

\subsection{Acoustic Sensing Module}
\begin{figure}[H]
\begin{center}
\includegraphics[width=.40\textwidth]{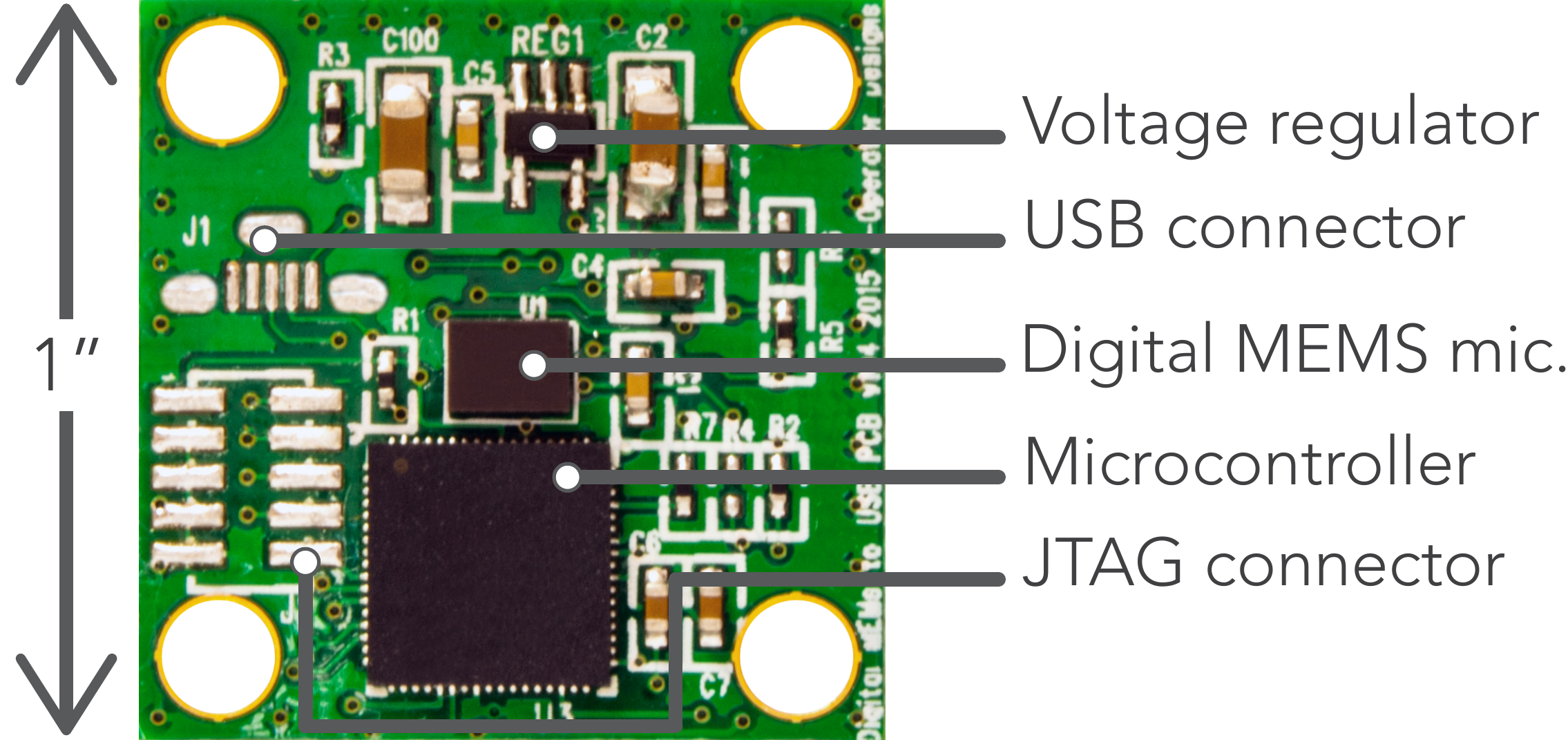}
\caption{Digital acoustic sensing module detail with main component annotations.}
\label{fig:edited_annotated_top}
\end{center}
\end{figure}

The sensing module (shown in more detail in Figure~\ref{fig:edited_annotated_top}) is mounted at the end of the gooseneck to the custom microphone mount and covered by the windshield for wind protection. The flexible gooseneck covers the microphone's USB cabling and allows for the positioning of the microphone for node mounting on horizontal or vertical surfaces such as window ledges or walls. The custom microphone mount provides a top hood for the microphone to reduce the chances of rain water dripping down the microphone modules front face and into the port. The mount's rear ports eliminate the chance of acoustic resonance within its internal cavity, which could color the microphone's signal at certain frequencies.

\vspace{-6pt}
The custom sensing module in Figure~\ref{fig:edited_annotated_top} is based around a digital microelectromechanical systems (MEMS) microphone. These were chosen for their low cost, consistency across units and size, which~can be 10$\times$ smaller than traditional devices. The model utilized here has an effective dynamic range of 32--120~dBA ensuring all urban sound pressure levels can be effectively monitored. It was calibrated using a precision grade sound-level meter as reference under low-noise, anechoic conditions, and~was empirically shown to produce sound pressure level (SPL) data at an accuracy compliant with the IEC~61672-1 Type-2 standard \cite{international2013electroacoustics} that is required by most US and national noise codes. This procedure is outlined for an earlier version of the sensor in \cite{Mydlarz_AppliedAcoustics17}. This digital microphone contains, within its shielded housing, an application-specific integrated circuit (ASIC) which performs the analog to digital conversion of the microphone's alternating current (AC) signal to a 1-bit pulse density modulated (PDM) digital signal. This~early stage conversion to the digital domain means there is the absolute minimum of low level analog audio signal moving around the circuit, resulting in superior external radio frequency interference (RFI) and localized electromagnetic interference (EMI) rejection. EMI from low-cost power supplies and the SBC are further reduced by the voltage regulator and array of capacitors, designed~to filter out any AC noise on the direct current (DC) input power rail. The PDM signal from the MEMS microphone is fed to the Microcontroller where it is converted to a pulse-code modulated signal (PCM), filtered~to compensate for the microphones frequency response and fed via USB audio (connector shown unpopulated in Figure~\ref{fig:edited_annotated_top}) to the master device, which in this case is our SBC. The enumeration of the sensing module as a USB audio device means it is SBC agnostic and so it has the potential to work with any USB enabled master SBC. The unpopulated JTAG connector allows the microcontroller to be flashed with updated firmware if required using an external programmer; however, on module power-up, it will briefly hold in a \textit{bootloader} mode allowing the firmware to be updated via USB. This~allows for future remote firmware updates over-the-air, administered by the SBC, via USB to the module's~microcontroller.

Before deployment, the component side of the sensing module is sealed with liquid electrical tape to ensure that resting water doesn't corrode the components or printed circuit board (PCB) traces. Silver traces were selected to reduce the chances of any corrosion spreading throughout the board. The PCB has extensive ground planes that run across each side for effective RFI and localized EMI shielding. The entire base of the microcontroller is soldered to this plane which acts as a heat-sink to spread its generated heat across the PCB. A positive side effect of this is the heating of the closely neighboring MEMS microphone. Whilst temperature variations are likely to have a minimal impact on the microphones sensitivity, it aids in maintaining a relatively constant temperature on the microphone diaphragm, reducing the effects of water condensation and the possibility of impaired operation in the event of water freezing anywhere near the microphone and forcing components out of place.

\subsection{Data Capture Module}
\begin{figure}[H]
\begin{center}
\includegraphics[width=0.6\textwidth]{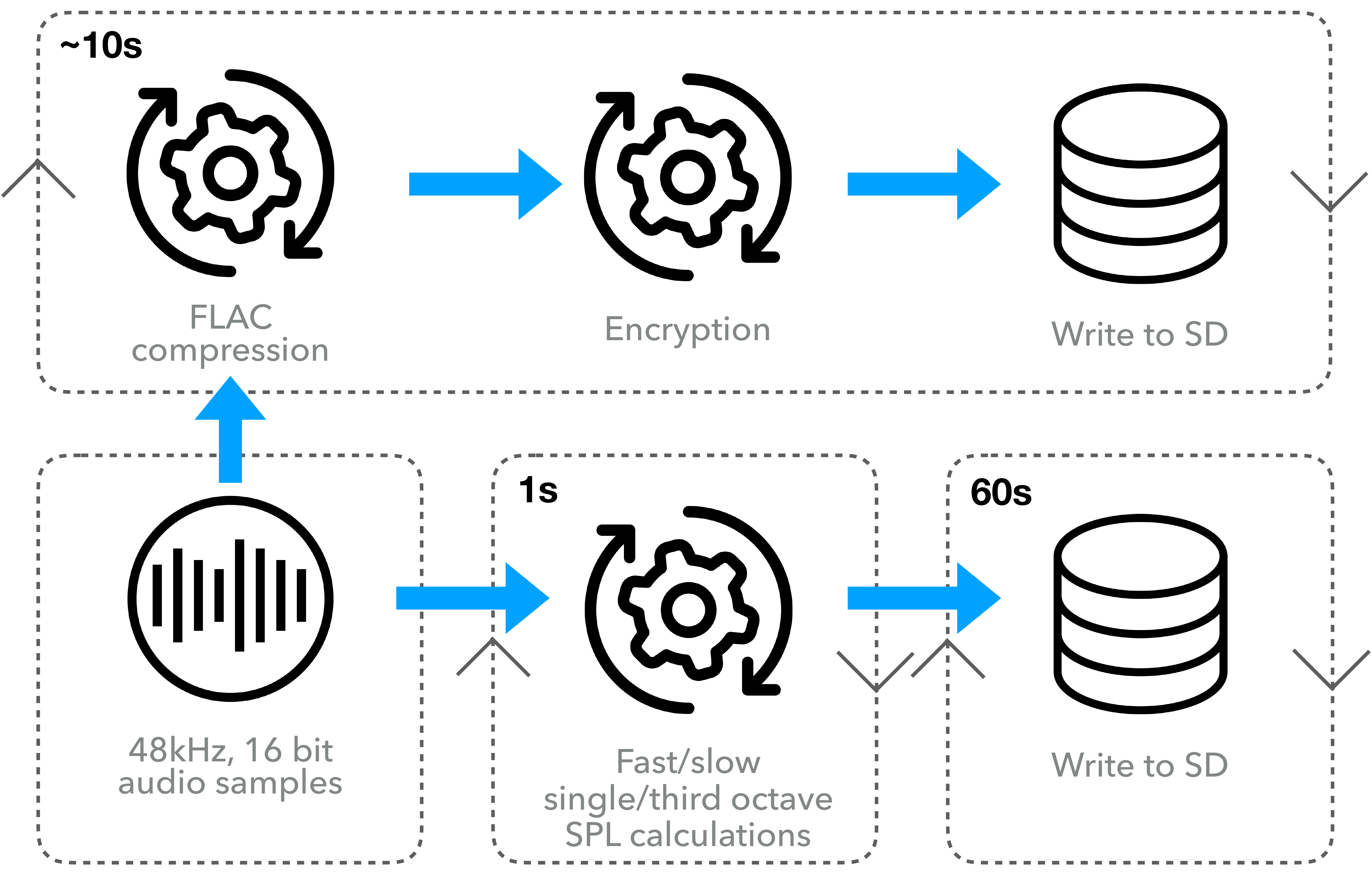}
\captionsetup{width=1\linewidth}\caption{Operational data flow of capture module showing the time-cycle of each procedure in top left.}
\label{fig:capture_flow}
\end{center}
\end{figure}

This module is responsible for: audio recording, compression, encryption, sound pressure level (SPL) extraction, and data writing to disk. It operates independently of all other sensor operations and has in-built monitoring for hardware level faults such as electrical microphone failure. The operational data flow is illustrated in Figure~\ref{fig:capture_flow}.

\vspace{-6pt}
The acoustic sensing module delivers 16~bit audio samples at 48~kHz via USB audio to the Raspberry Pi. These samples are continuously double buffered in 1~s chunks and handed off to either the SPL or audio operational processes. The accumulation of audio samples occurs in RAM for both processes to minimize secure digital (SD) card input/output (I/O) operations, which can lead to SD card fatigue and corruption. Single and third octave SPL data is written to a comma separated value (CSV) file in RAM every 125~ms and 1~s for the fast and slow integration times, respectively, including A, C and un-weighted SPL values. When 60~s worth of SPL data has been generated, it is written to a dedicated data partition on the SD card ready for data upload. Snippets of audio, 10~s in length are generated with random spacing between these 10~s samples to break up continuity of the audio samples for privacy reasons. When these samples have accumulated in RAM, they are compressed using the lossless FLAC \cite{flac_web} encoder to reduce file upload size. This FLAC file is then encrypted using AES encryption with its randomly generated AES password encrypted using key-based Rivest, Shamir, and Adleman (RSA) encryption. The encrypted FLAC file is stored with its encrypted password and transferred to the SD card for data upload. This secured audio data is only decryptable at the decryption server (described in Section~\ref{sec:data_access_viz}) where the RSA private key resides for enhanced data~security.

\subsection{Operational Modules}
\label{sec:op_func}
This collection of modules ensures the robust and autonomous operation of the sensor node and enables the adaptation to different adverse scenarios.

Networking modules: The networking module's goal is to keep the sensor securely and reliability connected to an internet connected network, via Wi-Fi, ethernet or cellular. A white-list of Wi-Fi access points is maintained on the sensor and the module will prioritize and dynamically connect to trusted networks broadcasting with higher Wi-Fi signal strength and quality, to ensure consistent connectivity and data throughput. This module also monitors for a specific priority Wi-Fi network called \textit{sonyc-lifeline}, which is broadcast by a portable Wi-Fi access point that the sensor team carries. When the sensor connects to this access point, it enables the sensor engineer to remotely connect to the sensors operating system (OS) via a secure shell (SSH) and perform maintenance tasks. This function is especially useful when troubleshooting difficult to reach sensors, such as those requiring ladders or lifts to access.

Uploader modules: These modules handle all data upload to the project ingestion servers (described in Section~\ref{sec:data_ingestion}). This includes audio, SPL and status data. Locally cached SPL and audio data will only be deleted when a successful upload response has been received from the ingestion server. Another uploader module monitors the SD card utilization and enacts a deletion policy when SD card usage reaches 95\%. Based on the data priorities mentioned in Section~\ref{sec:data_collected}, first audio data are deleted in a temporally interleaved fashion, and then SPL data. For example, at a 95\% SD utilization warning, the~oldest audio file will be deleted, at the next, the second oldest will be deleted. This policy ensures that optimal temporal data continuity is maintained in the event of loss of connectivity. The~12~GB data partition on the sensor allows for three~days of offline operation before the deletion policy is enacted. When the policy is in operation, SPL data can remain unaffected for around 14~days.

Monitoring modules: At the top level, a process control system manages the initialization and restarting of all modules in the event of a crash or system reboot. This includes monitoring CPU and RAM utilization of each module, which are restarted if certain thresholds are exceeded. To keep track of all operations, logging is handled by dedicated modules for each key sensor process, with all log files uploaded to the central server for future fault diagnosis. A sensor node's status variables are logged every 3~s and uploaded as described in Section~\ref{sec:data_collected} and investigated in Section~\ref{sec:status_analysis}.

\section{Sensor Network Infrastructure}
\label{sec:infrastructure}
\begin{figure}[H]
\begin{center}
\includegraphics[width=0.75\textwidth]{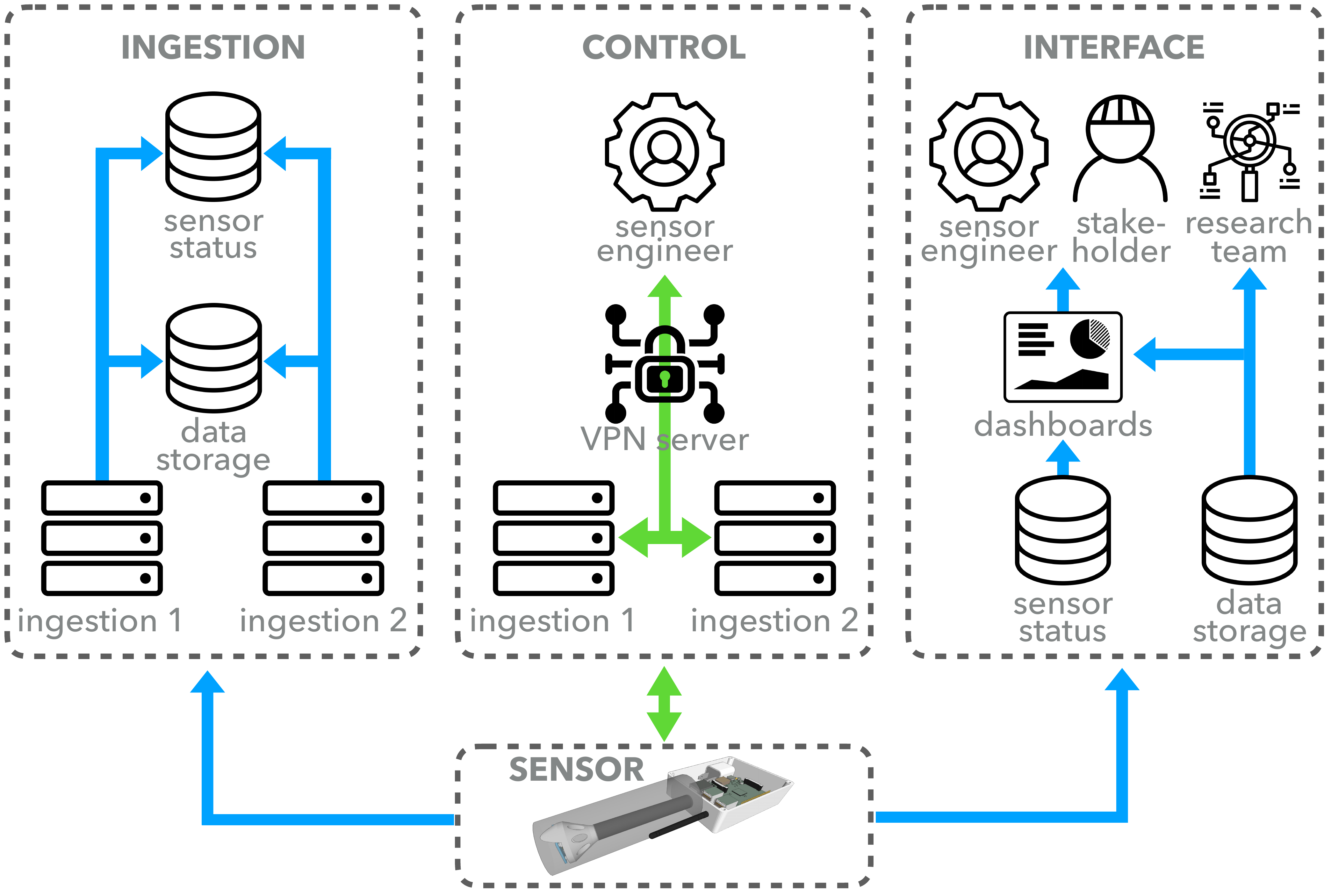}
\caption{Infrastructure overview showing core system operations with data flow in blue and control flow in green.}
\label{fig:infra_overview}
\end{center}
\end{figure}

Our data collection infrastructure consists of a number of physical and virtual servers handling: data ingestion, persistent data storage, secure data access, sensor control, data decryption, and sensor team/stakeholder data visualization via various dashboards. The following sections describe the major aspects of the infrastructure, with its focus on high uptime, scalability and data yield. Figure~\ref{fig:infra_overview}~gives a high level illustration of the core system operations and data flows that make up the network infrastructure and refer to Sections~\ref{sec:data_ingestion}, \ref{sec:control} and~\ref{sec:user_interface}.

\vspace{-16pt}
\subsection{Data Ingestion}
\label{sec:data_ingestion}

Two matched, high-power, physical servers make up the ingestion servers that act as the representational state transfer (REST) interfaces for all sensor data upload. Sensors are assigned to one of these independent servers to send data to when they establish a connection for load balancing, scalability, and high availability. Once data is uploaded to the ingestion server, it is cached to a local solid state drive (SSD), before it is securely transferred to the persistent network data storage drive in a data type/sensor/date folder structure. When a day's data folder is untouched for 24~h, it is compressed into a single tar archive for faster future access. The ingestion servers continue operating, albeit in a reduced state, if other components of the system fail. If the network data storage drive goes offline, the ingestion servers can cache data locally for several weeks.


\subsection{Remote Sensor Control}
\label{sec:control}
Each ingestion server runs two OpenVPN \cite{openvpn_web} based Virtual Private Network (VPN) services. Each VPN connects the ingestion servers to sensor nodes for direct sensor control via a dedicated control server. Each ingestion server has its own VPN address range that is automatically allocated to nodes as they connect. Sensor engineers can SSH directly into sensors when logged into the control server for remote troubleshooting and updating, regardless of which ingestion server the sensor is currently connected to. The sensor's code-base is maintained in a remote git repository cloned on the device, which can be remotely updated as a batch process, making network updates faster and more reliable than manually pulling changes.

\section{User Interface}
\label{sec:user_interface}
The SONYC sensor network currently has three main stakeholders who require access to its varying data types. The following sections describe the different user interfaces and functionalities offered to these stakeholders.

\subsection{Stakeholder Interface}
One of the project's key stakeholders is the NYC Department of Environmental Protection (DEP). This city agency is tasked, amongst other things, with enforcing the city's noise code \cite{noisecode_web} and is a key partner to the SONYC project. A prototype dashboard was developed that enables the enforcement schedulers to view aggregate sensor data in real time to identify possible noise hot-spots in the city.

\begin{figure}[H]
\begin{center}
\includegraphics[width=0.97\textwidth]{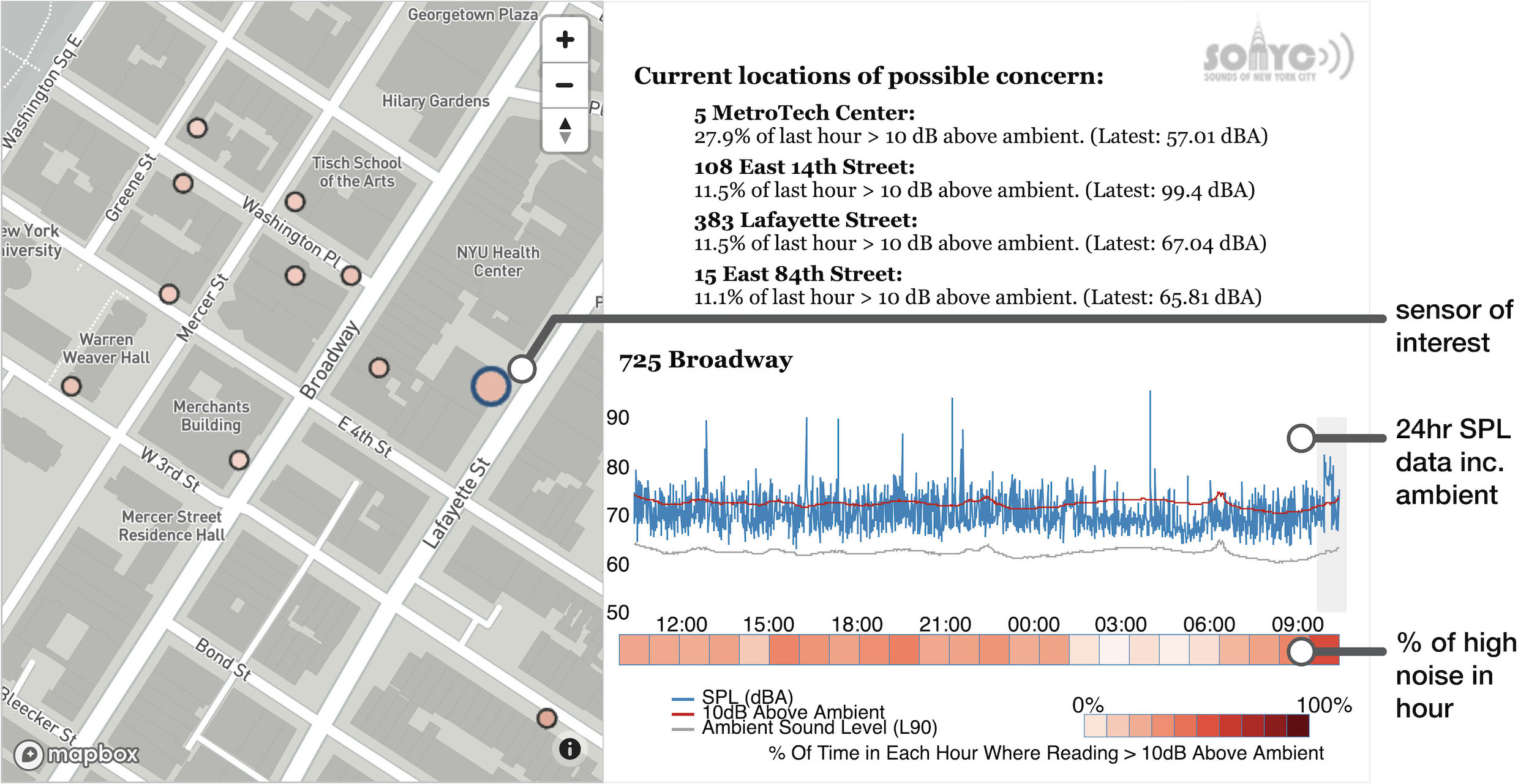}
\caption{Annotated prototype DEP dashboard showing real-time sensor data across a heavily deployed area of NYC.}
\label{fig:dep_dashboard}
\end{center}
\end{figure}

Figure~\ref{fig:dep_dashboard} shows the dashboard, which is being developed with the DEP to refine the data displayed and how it is aggregated and abstracted to provide useful and actionable information about urban noise. The map view allows the user to select a sensor of interest whose map icon color reflects the percentage of the last hour that the SPL exceeded 10~dB above ambient level. This percentage value and its associated sensor is shown in the upper right corner with an ordered list of \textit{locations of possible concern}, suggesting that excessive noise events are occurring in these locations. The plot below shows SPL data for the past 24~h and a corresponding heatmap below, colored on a scale representing the previously mentioned percentage of the hour in an excessively noisy state. This dashboard is being used alongside a DEP built alerting system that informs inspectors of when a localized cluster of noise complaints occur. Nearby sensor data are used to prioritize enforcement visits to those areas where excessive noise levels are detected for an extended period of time. Through collaborative iterations of this dashboard and alerting system, SONYC is integrating with the DEP's operational procedures for data-driven noise enforcement.

\vspace{-16pt}
\subsection{Data Access and Visualization}
\label{sec:data_access_viz}
Researchers looking to access SONYC SPL data can make use of a dedicated data access server which contains scripts with read-only access to the SPL data on the storage back-end. These scripts allow for data retrieval by: sensor, date range, data type, geographical ranges and data resolution. When complete, a data file per sensor is stored on the user's home directory in binary or CSV format.

A Noise Profiler tool was developed in-house that makes use of a custom Time Lattice data structure which allows for the querying of very large scale SPL data~\cite{Miranda:TimeLattice:CGF:18}. This web-based visualization framework supports geo-spatial and temporal sensor querying across large time ranges and offers insight at resolutions from years to seconds.

For the project's potentially sensitive audio data, a dedicated decryption server allows authenticated users access to the raw audio data. A user requires a valid signed certificate in order for the decryption server to return the AES password needed to decrypt an audio file. This ensures that anyone accessing this audio data has been vetted and cleared by the SONYC admin team as part of the certificate generation process.

\subsection{Sensor Network Monitoring}
\label{sec:monitoring}
The ingestion servers run a distributed Elasticsearch~\cite{divya2013elasticsearch} database which stores and indexes sensor telemetry updates. This database allows our sensing team to obtain a near real-time and, if needed, historic view of the entire sensor network's operational status. A sensor node's static meta-data is stored in a MySQL database, interfaced using a headless Content Management System (CMS), which~provides a web interface for managing sensor information, as well as a REST application programming interface (API) for programmatic access. This~meta-data includes the name of the sensor or location deployed, date deployed, life stage (Active,~Testing, Retired, etc.), geographic location, and engineer notes. These~notes contain details about sensor changes made on maintenance visits, details about how and where it was mounted, and any idiosyncrasies about that specific node or deployment that would be relevant to the sensor engineer who needs to provide maintenance.

\begin{figure}[H]
\begin{center}
\includegraphics[width=0.97\textwidth]{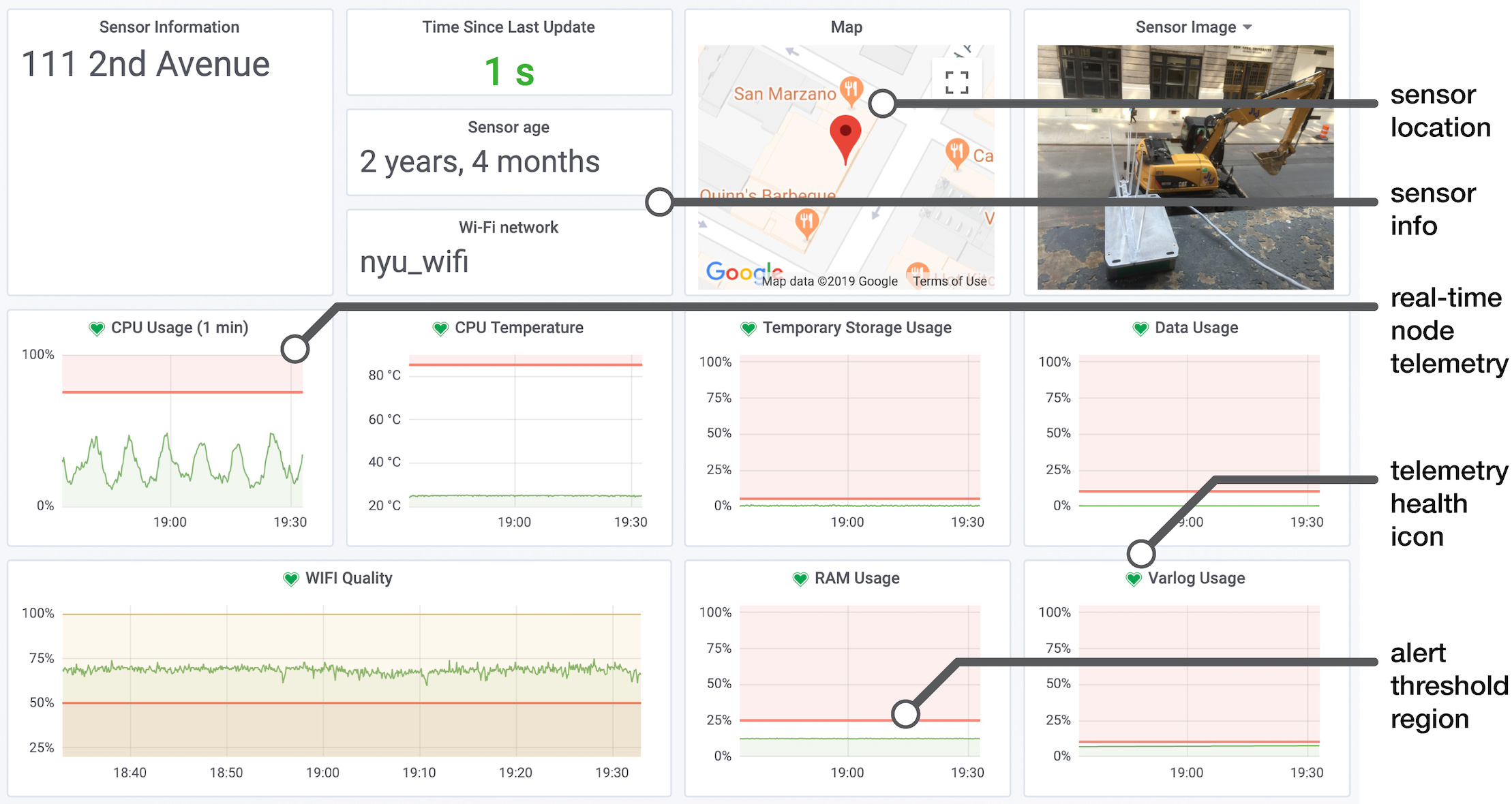}
\caption{Annotated sensor dashboard showing dedicated real-time sensor information.}
\label{fig:sensor_dashboard}
\end{center}
\end{figure}

In order to maintain network uptime, it is necessary that we are able to oversee the health of the network. A main dashboard was developed to provide at a glance summary info, which can display the current status of all nodes on a single page.
This dashboard displays a table of summary statistics describing the overall health of the network, a map displaying all active sensor node deployment locations, and a sortable, filterable table listing each node and relevant information about the node's health such as time since last update, Wi-Fi quality, data usage, etc. Each sensor entry has a link to enable editing of a sensor's meta-data and a link to a more detailed and dedicated dashboard for each sensor shown in Figure~\ref{fig:sensor_dashboard}. This dedicated dashboard is rendered using Grafana \cite{grafana_web}, an open-source data visualization platform specializing in time-series data. It displays a grid of plots detailing the node's operation over time, including Wi-Fi strength and quality, CPU usage, CPU~temperature,~etc.

\vspace{-6pt}
An important feature of Grafana is the ability to set \textit{alerts}, which are depicted by the red lines and red shaded regions of the plots in Figure~\ref{fig:sensor_dashboard}. This provides an interface to setup when alert notifications occur based on custom thresholds and rules, such as: RAM usage exceeding 25\% for longer than 10~min. These notifications can be directed to many different platforms. In our case, a sensor's alert is sent to a dedicated sensor engineer alerting Slack \cite{slack_web} messaging app channel and to the smartphone messaging app, Telegram \cite{telegram_web}. The green hearts display the current health status of a particular telemetry parameter. The time history of this state is also logged for retrospective fault diagnosis. To explore the challenges and lessons learned in maintaining the SONYC sensor network, the following section covers the analysis of the network's downtime.

\section{Sensor Network Downtime Analysis}
Maintenance of the sensor network relies on speed, as a fault will typically begin to manifest and, within a few hours or days, result in a sensor going offline, requiring physical on-site intervention. This~reactive approach is costly and can usually be avoided if foresight of an impending failure is given to the sensor engineer team, where a fix can be done remotely. Real-time monitoring and alerting provide this foresight and is critical for the successful upkeep of a sensor network at any scale, especially~those with hard to reach sensor nodes where access is costly and time consuming. A more predictive approach to sensor maintenance allows for forward planning and faster response times to impending faults. This includes the ability to inform sensor engineers on predicted fault types before a repair is made, reducing the involvement of more costly senior engineers in the maintenance process.

To explore the operational conditions of the sensor network, an analysis was carried out that focuses on downtime through its data yield, which will be defined in Section~\ref{sec:data_yield}. This analysis makes use of the sensor status or telemetry data collected during an 18-month period, as detailed in Section~\ref{sec:status_analysis}.

\subsection{Data Yield}
\label{sec:data_yield}
In its current iteration, the sole purpose of each SONYC sensor node is the delivery of usable noise data to the project's ingestion servers. In order to define data yield, we decided to focus on the node's ability to successfully and continuously upload one-minute sound pressure level (SPL) files, as~this process occurs continuously throughout the node's operational life. A 100\% data yield would therefore mean 60 readable/uncorrupted files are present on the storage servers for every operational hour. To~explore this, the data yield was calculated per hour for an 18-month period between January~2017 and July 2018 for nodes that were deployed on or before 1 January 2017, so they were expected to be operational for the entire period.

\begin{figure}[H]
\begin{center}
\includegraphics[width=0.97\textwidth]{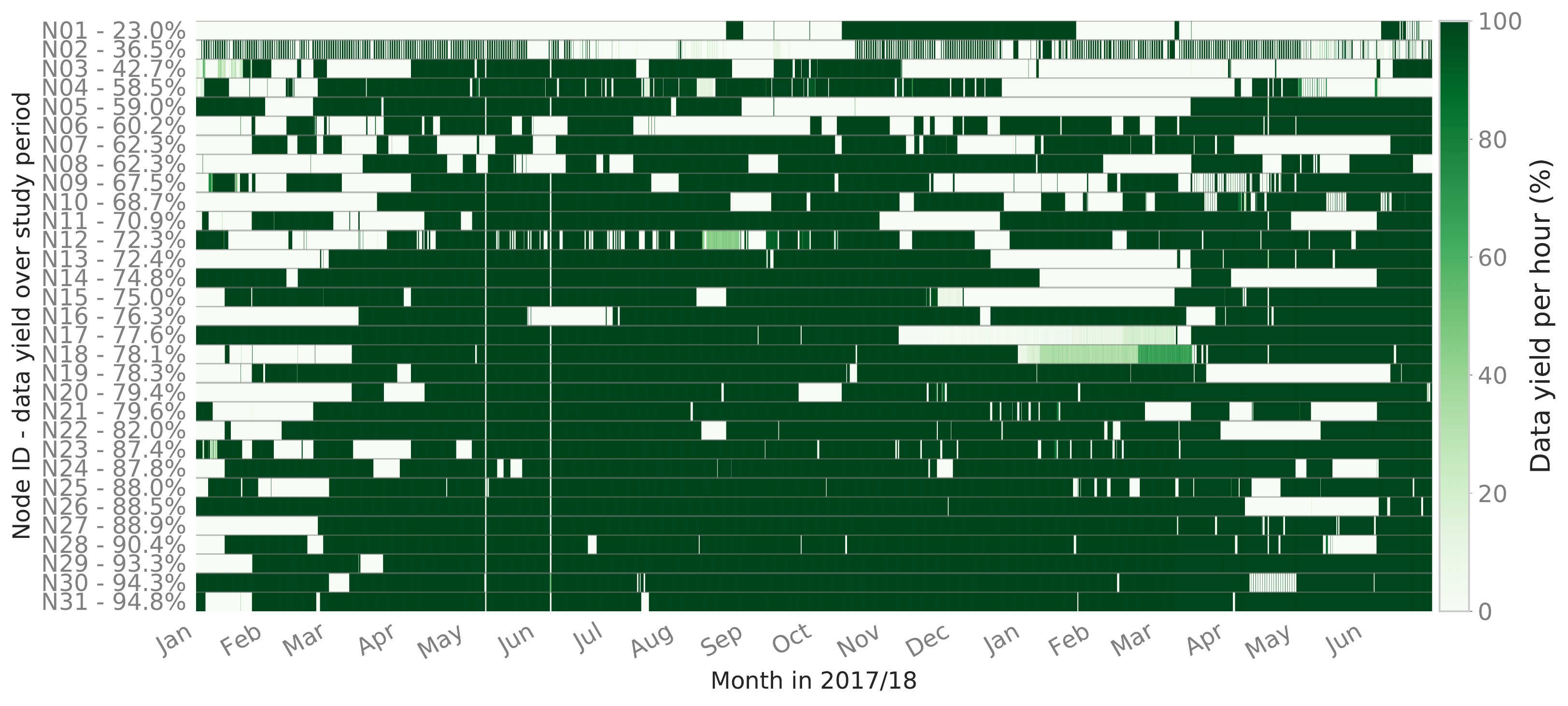}
\caption{Sensor network data yield percentage by hour (January 2017--July 2018).}
\label{fig:network_yield}
\end{center}
\end{figure}

Figure~\ref{fig:network_yield} visualizes this data yield throughout the study period for 31 nodes deployed across New York City (NYC), labelled N01-N31, with the total data yield percentage per sensor shown. The~intensity of the color represents the data yield as a percentage for each hour, where the nodes are ordered by increasing data yield. Node yield varies significantly with the lowest performer successfully transferring 23.0\% of its data up to the highest performer at 94.8\%, with a mean yield of 73.2\% and median of 76.3\%. Interestingly, the total deployment duration or \textit{age} of a sensor does not relate to its propensity for downtime. Based on observations of the sensors in operation, N02, for example, has had issues with its Wi-Fi access point nearby losing power sporadically, which produces this patchy data yield pattern. Larger blocks of downtime such as those seen on sensors N13--17 over the winter of 2017/18 were partly due to power outages caused by cable damage from window closures on the cables and power supply damage by users of the spaces where sensors are deployed. Earlier in the study period, it can be seen that the network as a whole shows more downtime. This is likely due to code bugs that were fixed in the Spring of 2017, shown by the general uptime improvements after April~2017. Sensor team staffing also increased around this time and the repair procedures were optimized, resulting in faster response times to downed sensors and thus increased data yield.

These difficult to diagnose downtime periods could have occurred for a number of reasons such as: sensor engineer staffing and/or technical issues. To begin to explore this, the following section explores the analysis of the sensor network's telemetry data with an aim to understand more about the possible cause of these downtime periods and ways in which they can be predicted.

\vspace{-16pt}

\subsection{Telemetry Analysis and Fault Diagnosis}
\label{sec:status_analysis}

The telemetry data collected every 3~s over the 18-month period between 1 January 2017 and 30 June 2018 from each deployed sensor node provide an insight into the operational status of the sensor network. Across the 31 sensors under study, this consists of 462.1~million rows of telemetry data, across 10 variables:
\begin{itemize}[leftmargin=*,labelsep=5mm]
\begin{minipage}{\textwidth} \item\textbf{CPU load (\%/1 min):
} mean CPU usage over a 1~min period across all four 900~MHz CPU cores;
\end{minipage}
\begin{minipage}{\textwidth} \item\textbf{CPU load (\%/15 min):} mean CPU usage over a 15~min period across all four 900~MHz CPU cores\end{minipage}
\begin{minipage}{\textwidth} \item\textbf{CPU temp ($^{\circ}$C):} core CPU temperature in degrees Celsius;\end{minipage}
\begin{minipage}{\textwidth} \item\textbf{RAM usage (\%):} usage of 925~MB RAM;\end{minipage}
\begin{minipage}{\textwidth} \item\textbf{Wi-Fi signal strength (\%):} measure of Wi-Fi signal strength;\end{minipage}
\begin{minipage}{\textwidth} \item\textbf{Wi-Fi signal quality (\%):} measure that factors in signal-to-noise ratio (SNR) and signal strength;\end{minipage}
\begin{minipage}{\textwidth} \item\textbf{Data usage (\%):} usage of 12~GB SD card data partition;\end{minipage}
\begin{minipage}{\textwidth} \item\textbf{TMP usage (\%):} usage of 50~MB RAM disk partition used for fast temporary I/O operations;\end{minipage}
\begin{minipage}{\textwidth} \item\textbf{Var-log usage (\%):} usage of 50~MB RAM disk log partition where all log files are written to;\end{minipage}
\begin{minipage}{\textwidth} \item\textbf{Running processes:} count of running processes.\end{minipage}
\end{itemize}

To explore the condition of the network just prior to a period of downtime, an hour long \textit{prefail} section of telemetry data was selected just prior to each period of downtime greater than 6~h in length. A \textit{stable} section is defined as that which has at least 48~hours of 100\% data yield/uptime before and after it.

\begin{figure}[H]
\begin{center}
\includegraphics[width=0.6\textwidth]{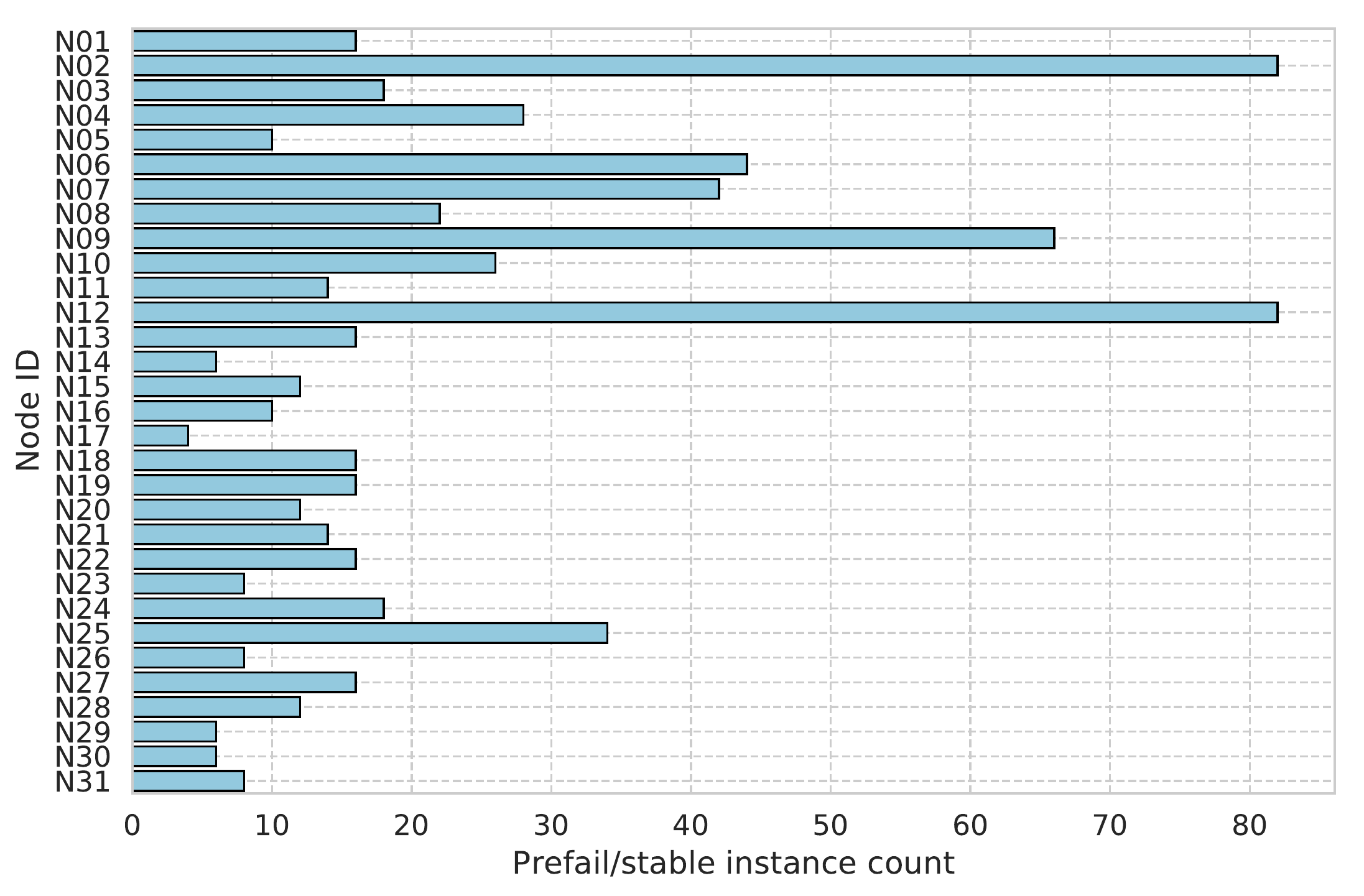}
\caption{Prefail/stable instance count by sensor.}
\label{fig:sensor_inst_count_bar}
\end{center}
\end{figure}

An equal number of prefail/stable sections were selected from each sensor. This results in a total of 688 prefail/stable instances with 344 in each class and a mean count of 22 total per sensor. Figure~\ref{fig:sensor_inst_count_bar} shows the total prefail/stable instance counts per sensor. The~prefail class contains 241,803 observations with the stable containing 260,047. This discrepancy is due to the low priority and lack of any local caching of the telemetry data on the sensor. It is likely that, during a prefail period, these data were unsuccessfully transmitted, in which case it is discarded and a new set of observations is sent a few seconds later.

With this labelled prefail/stable telemetry data-set, the distributions were plotted for each variable to uncover any potential differences in their means as a first step in discovering separation between the two classes. These data can be heavily skewed in some instances. Percentage variables are particularly susceptible to this, such as in the cases of \textit{disk partition usage} and \textit{Wi-Fi signal} fields. Under normal/stable sensor operation, these tend to remain at very low or very high values, respectively. During unstable periods, such as weak Wi-Fi conditions, they can manifest at the opposite extremes, resulting in these heavily skewed distributions. For example, under normal operation, the temporary file system (TMP) disk partition should remain at $\approx$0.1\%. This increases and holds at 100\% under particular fault scenarios.

\begin{figure}[H]
\begin{center}
\includegraphics[width=0.97\textwidth]{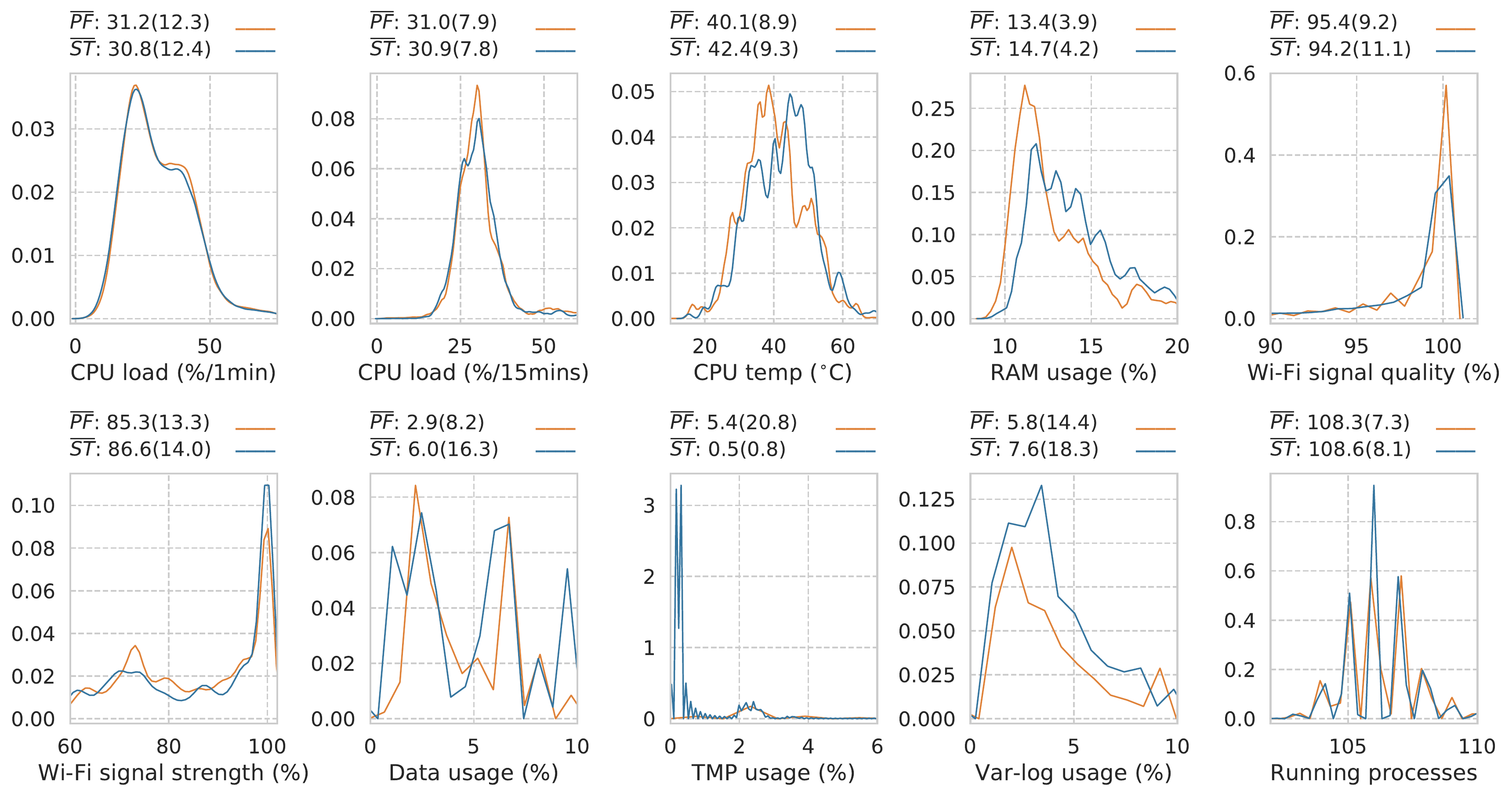}
\caption{Normalized prefail/stable distributions across telemetry variables including mean and standard deviation in brackets ($\overline{PF}$ = prefail mean \& $\overline{ST}$ = stable mean). z-test $\rho$ all \textless 0.001 (prefail~$n$~=~241,803 and stable $n$~=~260,047).}
\label{fig:hist_prefail_stable_all_vars}
\end{center}
\end{figure}

\vspace{-6pt}
Figure~\ref{fig:hist_prefail_stable_all_vars} reveals significantly different means between each group for all telemetry variables using a comparison of means z-test with $\rho$-values all at the \textless0.001 level. CPU load (\%/1min): z~=~$-$19.52; CPU load (\%/15 min): z~=~$-$26.02; CPU temp ($^{\circ}$C): z~=~108.65; RAM usage (\%): z~=~138.83; Wi-Fi signal quality (\%): z~=~$-$31.95; Wi-Fi signal strength (\%): z~=~37.44; Data usage (\%): z~=~67.97; TMP usage (\%): z~=~$-$120.87; Var-log usage (\%): z~=~22.21; and running processes: z~=~33.54. The size of this data-set ensures that the mean of each classes samples will be approximately normally distributed, making the z-test suitable in this case.

\vspace{-6pt}
Whilst showing significant overlap within each class's distribution, the significantly different means reveal a level of separation worthy of further investigation. In terms of RAM usage, the stable instances show an increase in resource usage suggesting that, under normal operating conditions, the sensor is doing more. The variation in this variable as shown by the increased standard deviation value of 4.2 under stable conditions versus 3.9 under prefail conditions suggests more dynamic sensor behaviour, indicative of normal operation. Wi-Fi signal quality, counter-intuitively, shows a slightly lower mean value under the stable class as lower signal quality would typically result in data upload difficulties. However, this slight shift would not generally result in these issues. Signal strength shows a higher value in the stable condition, which is as expected. The disk usage variables \textit{TMP} and \textit{Var-log} show expected mean values as the TMP folder should not exceed 1\% usage under stable operation and Var-log typically accumulates log files faster if all scripts are operating normally and carrying out their normal logging procedures.

\begin{figure}[H]
\begin{center}
\includegraphics[width=0.6\textwidth]{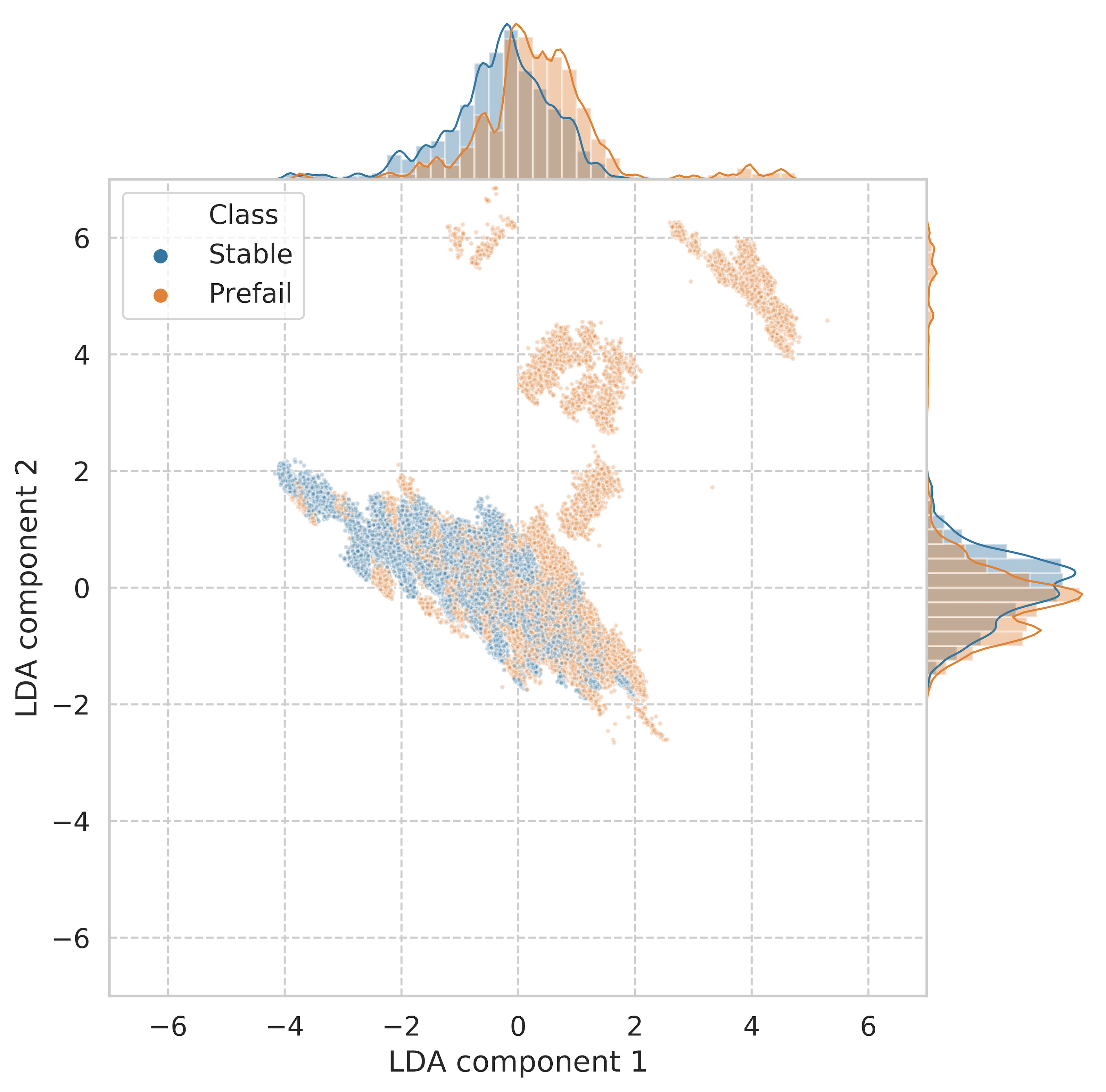}
\caption{First two components of linear discriminant analysis showing prefail/stable condition separation including their distributions.}
\label{fig:lda_2d_comps}
\end{center}
\end{figure}

To delve deeper into the apparent separation between the prefail/stable classes with respect to these telemetry variables, a linear discriminant analysis (LDA) was used to visualize this separation in a multidimensional space. LDA is a dimensionality reduction and visualization technique that aims to uncover linear combinations of variables that best explain the separation between a number of classes. The 10 telemetry variables were fed into the LDA, with two components extracted for visualization. Figure~\ref{fig:lda_2d_comps} shows these two output components in a scatter plot, with distinct regions of separation between the prefail and stable classes, mainly within component 1. The stable and prefail means were shown to be significantly different using an independent samples \emph{t}-test within component 1 (stable mean = $-$0.330 and std = 0.914, prefail mean = 0.355 and std = 1.085). This~suggests that the prefail/stable states can be differentiated between using statistical approaches based on the collected telemetry data. This separation is particularly pronounced when observing the data in a higher dimensionality space.

\vspace{-6pt}
To begin the translation of this apparent prefail/stable separation into a practical predictive model for sensor fault prediction, a random forest with 1000 estimators was trained on a stratified 80\% random split of the data. This train/test split was performed across prefail/stable instances rather than the raw telemetry data entries, in order to ensure that a balanced number of prefail/stable instances were present within each split. After this splitting process, the test data were standardized by removing the mean and scaling to unit variance. The same scaling procedure that was fit to the test data was used to scale the training data to ensure any subsequent model has no prior knowledge of the test-sets' distribution. Table~\ref{tab:data_split} describes the counts and percentages within each split group and then within each respective class. This reveals a relatively balanced instance and row split between prefail/stable classes within each of the train/test splits.

\begin{table}[H]
\centering
\caption{Train/test data split description showing prefail/stable instance row counts and percentages.}
\label{tab:data_split}
\begin{tabular}{ccccc}
\toprule
\bf{Split Group}                            & \bf{Total Instances (Rows)}       & \bf{Class}    & \bf{Instance $n$ (\% of Total)}         & \bf{Row Count (\% of Total)}  \\
\midrule
\multirow{2}*{Train}                        & \multirow{2}*{550 (399,959)}      & Stable	    & 276 (50.2)			                & 204,304 (51.1)			    \\
&                                   & Prefail	    & 274 (49.8)			                & 195,655 (48.9)		        \\

\multirow{2}*{Test}                         & \multirow{2}*{138 (101,891)}      & Stable	    & 68 (49.3)			                    & 55,743 (54.7)			        \\
&                                   & Prefail	    & 70 (50.7)			                    & 46,148 (45.3)		            \\\bottomrule
\end{tabular}
\end{table}

\begin{figure}[H]
\begin{center}
\includegraphics[width=0.8\textwidth]{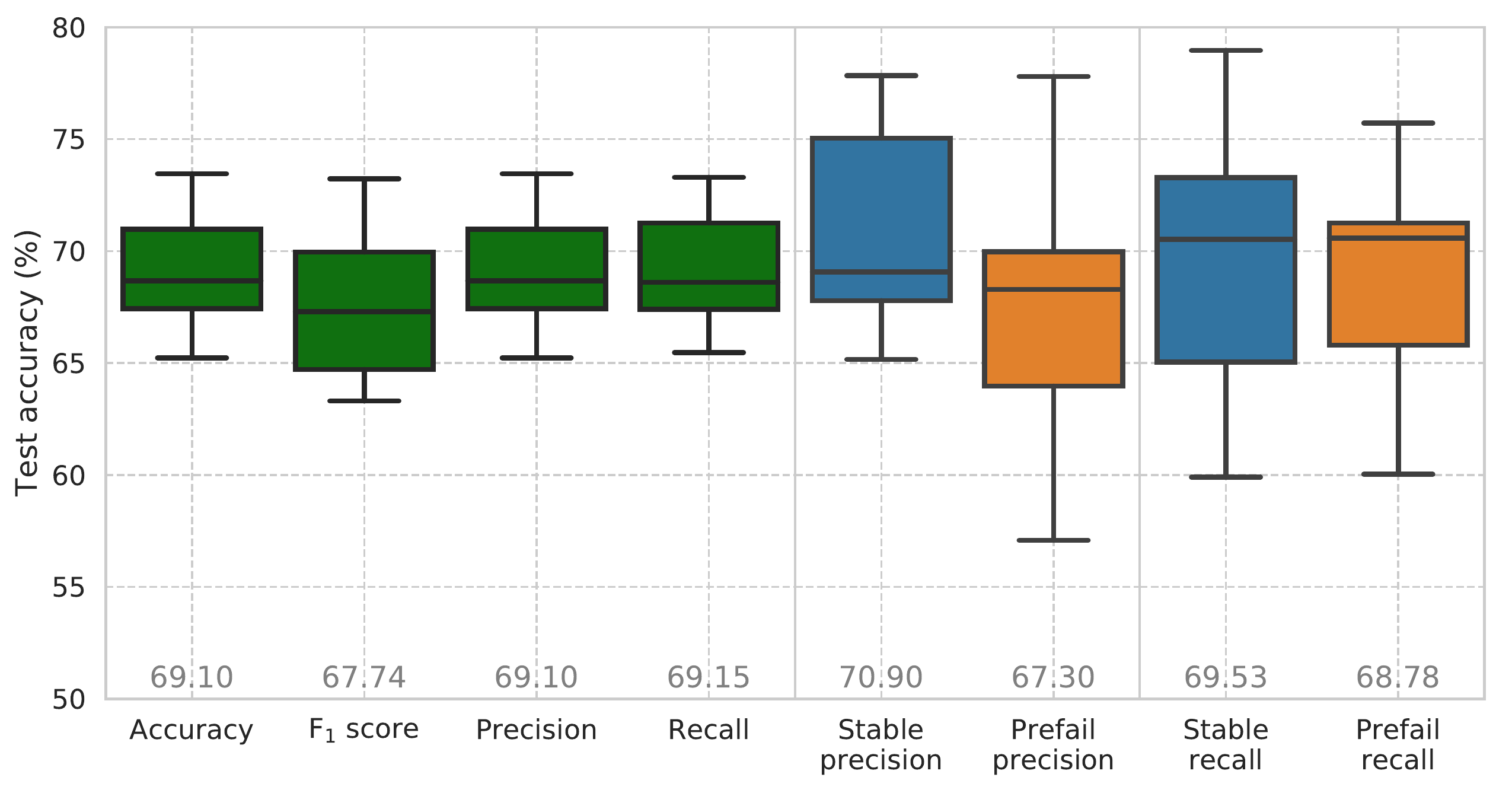}
\caption{Average and class accuracies including precision--recall scores for random forest model (10~trials  with randomized initialization and shuffling of the training and test data, mean values displayed at plot bottom).}
\label{fig:box_plot_rf_results}
\end{center}
\end{figure}

To compensate for statistical fluctuations, 10 models were trained with randomized initialization and shuffling of the training and test data. Figure~\ref{fig:box_plot_rf_results} shows the results of the model's use on the test data-set as box plots. Average precision and recall values of 69.10\% and 69.15\%, respectively, show the models relatively high predictive abilities. The model shows an increased ability at correctly identifying stable conditions with a class precision value of 70.90\%. This could be exploited in a subsequent implementation by raising a flag when a \textit{non-stable} state is detected by the model. Increased recall scores in the stable class suggest that the model is slightly better at avoiding the false detection of stable running, which would be a prefail state. Again, in practice, a flag could be raised when a \textit{non-stable} instance is detected.

\vspace{-6pt}
In the application of sensor fault prediction, it is important to minimize the occurrence of false negative errors where a sensor is in fact moving into a failure state and the model fails to predict that this is happening. An undetected sensor failing results in the need for a costly physical intervention. A~false positive prediction of impending sensor failure when in fact the sensor operating in a stable state is not ideal but certainly carries with it less of the costly ramifications of a false negative prediction. The time required to remotely check on a falsely failing flagged sensor is far less than physically visiting a downed node. Therefore, when choosing a threshold at which a sensor is considered to be in a prefail state, a lower value is preferred in this case to ensure that all potential prefail instances are flagged. To~this end, a high recall score is more important in this application.

\begin{figure}[H]
\begin{center}
\includegraphics[width=0.6\textwidth]{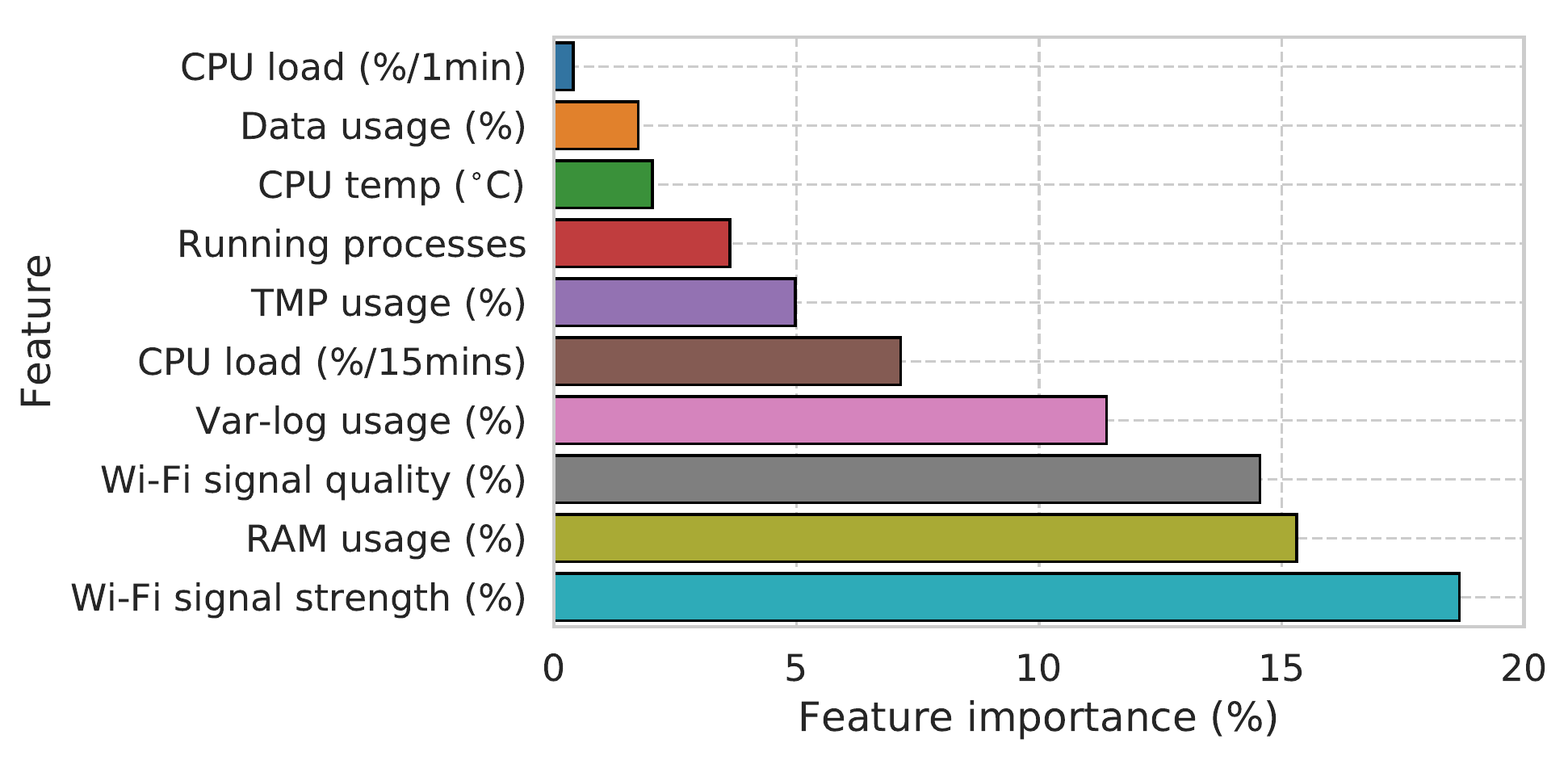}
\caption{Feature importance's of predictive model (total = 100\%).}
\label{fig:feat_importance}
\end{center}
\end{figure}

The averaged feature importance of the model shown in Figure~\ref{fig:feat_importance} reveals the influence of Wi-Fi signal characteristics, RAM usage, and Var-log usage on prefail state. This also reveals that this initial model is picking up on real-world failure modes. It should be noted that observations of poor Wi-Fi conditions are common and have been seen to be one of the main causes of sensor downtime. Node~N02, for example, exhibits a weak Wi-Fi signal shown clearly by its inability to continuously upload data from January to July in Figure~\ref{fig:network_yield}. The more sparse data yield periods between July and October were caused by the outage of the nearby Wi-Fi access point, which caused the node to switch to a more remote access point, further reducing its Wi-Fi signal strength and ability to upload its data payload over time. In response to this, high-gain directional Wi-Fi antenna have been deployed to improve signal strength on nodes struggling with this issue.

\vspace{-6pt}
RAM and Var-log usage variables are also intuitively linked to sensor condition as abnormal values of these suggests operational scripts are not operating correctly. Script faults can produce a number of complex symptoms that affect node functionality, including: memory leaks, connectivity issues, and total script crashes. A notable fault observed resulted in the failure of the SPL logging code to reset the CSV file causing its uninhibited growth on the TMP folder (captured by the TMP usage telemetry variable). This partition is limited to 50~MB in size. When this partition fills to 100\%, all~subsequent audio file write-to-disk operations fail and the device effectively stops collecting data. This process takes around three~days to enter a fully failed state where the TMP partition is filled. The~node does not lose its connectivity to the network but cannot generate data.

In addition to these aforementioned faults that typically develop over some period of time, there are more instantaneous faults that result in data loss. The first of these is power failure, caused by an internal electrical failure of the node or power cable removal by a third party. This is usually localized to particular sensors. To mitigate this, in locations where the node is plugged into an accessible outlet, a sealed cover is applied to the power outlet in use. The second instantaneous fault type is a server level fault where there is a loss of small chunks of data caused by a server fault/outage. This type of fault is rare but typically has network wide ramifications. All nodes experienced data loss for two~days in May and June 2017 when a server level data move procedure failed.

\subsection{Summary}
The data yield or uptime of the SONYC sensor network varies widely between sensors. These~periods of downtime can be caused by a number of known and unknown fault conditions. Faults that result in telemetry data changes prior to a period of sensor downtime provide, if they can be detected, a window of opportunity for a sensor engineer to make a remote fix, saving a costly physical on-site repair. To this end, a predictive model was developed that made use of this real-time telemetry data to determine if a sensor was moving into a \textit{pre-failure} state. The presented random forest model for pre-failure detection shows promising scores for accuracy and precision--recall. With this ability to flag prefail periods in real time from the sensors telemetry data, a system can be implemented to alert a sensor engineer who could make a remote fix to the sensor before a period of downtime occurs.

Due to the variable data yield of each sensor node, some exhibited more/less examples of prefail periods so the instance sampling is not equal across all sensors. This can be seen in the high prefail/stable instance counts for sensor nodes N02, N09, and N12, shown in Figure~\ref{fig:sensor_inst_count_bar}. These~presented models may be influenced by this and could mean that they are biased to commonly failing sensors and their particular fault states. With larger data-sets, the between-sensor balance can be brought back into line, eliminating this potential bias.

One reason why the model may not be showing higher accuracy scores is the fact that it is predicting on a wide range of fault conditions that produce varying telemetry data changes. As noted in the previous section, there are instances where a sensor may go down because of a power failure at the sensor or at the Wi-Fi access point the sensor is connected to. In this case, the sensor's telemetry data at its prefail period (1~h prior to its downtime) would presumably exhibit \textit{stable} characteristics as the sensor is functioning perfectly well. In these instances, it would be unlikely that the model would provide any indication of the sensor moving into a prefail state.

\section{Conclusions}
We have presented a large scale noise monitoring sensor network consisting of over 55 sensors with advanced edge compute capabilities. This network has been operational for close to three years and has amassed a large amount of SPL and audio data. The upkeep of a sensor network of this size is a challenge as sensors go down for a number of reasons. Foresight of sensor failure can avoid the need for costly and time-consuming on-site physical intervention. The telemetry data generated from each sensor node has proven to contain vital information on a sensors' health that has enabled the development of a prototype predictive model for fault detection. With confident predictions of impending failure, a sensor network can become more resilient and ultimately generate more valid data. The pipelines outlined in this paper aim to increase this uptime and data generation for distributed sensing applications where sensor access can be difficult or costly. With the proliferation of Internet of Things (IoT) remote sensing applications, these approaches and studies of uptime and data yield become more relevant. The future work to advance this prototype system will be addressed in the following section.

\section{Future Work}
\label{sec:future_work}
The next stage in the development of the fault prediction pipeline is to implement it as a real-time model acting on data arriving from live sensors in the field. Any flags raised over time can be fed into the Elasticsearch database discussed in Section~\ref{sec:monitoring}, which allows for the real-time monitoring of sensor condition via a Grafana dashboard. Prefail flags can be hooked into the existing sensor engineer alerting system also discussed in Section~\ref{sec:monitoring} so engineers can respond with haste.

With more telemetry data collection, a more balanced data-set can be created that contains equal numbers of prefail conditions across the entire sensor network, reducing the models' bias towards a particular sensors issues. A full analysis of its ability to generalize to new or untrained on sensor data will also be performed to assess its practical use in condition monitoring.

It could be argued that a prefail period of one~hour does not afford a sensor engineer enough time to make a remote repair. To investigate this, one~hour prefail periods will be sampled with increasing time deltas from the sensor downtime period and observing the models accuracies as this increases. The~point in time at which the model's accuracy (with a focus on its recall values) falls~to an unacceptable level is the amount of prefail foresight the system can reliably provide. The use of a more advanced predictive model beyond a random forest may also be able to provide a more nuanced detection of \textit{pre-failure} states, including more fore-warning.

A useful and practical addition to the model would be the ability to predict failure type so that a sensor engineer can prepare for a repair ahead of time. A multi-class model could be trained that returns a prefail flag with a specific fault type label. This, however, would require the generation of a labelled data-set of different fault types, which currently doesn't exist. One way to begin the collection of this would be for sensor engineers to log repairs and categorize fault conditions as they are discovered in the field.

Another notable fault condition not previously discussed is a microphone failure where distortion is present on the audio signal due to an electrical or mechanical microphone fault. When this is occurring, the SPL and audio data generated is not trustworthy, so is therefore an important consideration. Some work has begun on training machine learning models that are able to detect this when run over the audio samples from the microphone. These models would be run at the sensor network edge with their predictions transmitted up as telemetry data. This would also be integrated into the fault detection pipeline with the possibility of sensor engineer alerting. All of these additions should result in an increased data yield from the sensor network.

The sensor presented in this article allows for the accurate, continuous monitoring of sound levels across a city. Whilst the gathering of accurate SPL data in situ is crucial to the monitoring of noise in smart cities, identifying the source of these noise events is of great importance. The sensor's powerful processing unit means that there is the capability of performing additional analysis of the audio signal. In tandem with the sensor development, considerable efforts have been employed on machine listening algorithms for the automatic identification of urban sound sources \cite{salamon2017ieee,Salamon:UnsupervisedUrban:ICASSP:15,Salamon:ScattteringUrban:EUSIPCO:15}. In its current configuration, the sensor collects 10~s audio snippets for manual sound source \textit{tagging/labelling}~\cite{cartwright2019crowdsourcing}. These labelled audio snippets are used as training data for the development of these algorithms. One of the key advantages of running these classification models directly on the sensing device is that there is no need to transmit audio data to a centralized server for further analysis, in this way abating possible security and privacy concerns related to the recording of audio data. However, porting these models to the device presents a challenge due to the models' high computational complexity. To this end, future work will involve research into \emph{model compression} \cite{Ba:ReallyDeep:NIPS:14}, which can be used to obtain the performance of deep learning architectures using shallow ones that require less computational~resources.


\vspace{6pt}

\authorcontributions{C.M., M.S., Y.L., B.S., and C.S. designed and built the SONYC infrastructure including sensors and all associated infrastructure; C.M. and Y.L. carried out the experimental data gathering; C.M., Y.L., and J.P.B. conceived and designed the experiments; C.M. performed the experiments; C.M. analyzed the data; C.M. wrote the paper.}

\funding{This work was funded in part by The United States National Science Foundation (NSF award CNS-1544753) and an initial seed fund from the New York University, Center for Urban Science and Progress~(CUSP).}
\acknowledgments{This work was supported in part by: The Sounds of New York City (SONYC) project (NSF~award CNS-1544753). The authors would like to thank the New York University, Center for Urban Science and Progress (CUSP) for their initial seed funding and ongoing support for the project, The Cooper Union for the use of their acoustic laboratory and the Department of Environmental Protection (DEP) for their input and ongoing support.}


\conflictsofinterest{The authors declare no conflict of interest. The founding sponsors had no role in the design of the study; in the collection, analyses, or interpretation of data; in the writing of the manuscript, and in the decision to publish the results.}



\reftitle{References}

\end{document}